\documentclass[reprint, amsmath,amssymb, aps,floatfix]{revtex4-2}

\usepackage{graphicx}
\usepackage{dcolumn}
\usepackage{bm}
\usepackage{hyperref}
\usepackage[mathlines]{lineno}
\usepackage{float}
\usepackage{appendix}
\usepackage{microtype}
\usepackage[all]{xy}

\begin{document}


\title{Overcome Competitive Exclusion in Ecosystems}

\author{Xin Wang}
\affiliation{Channing Division of Network Medicine, Brigham and
Women's Hospital and Harvard Medical School, Boston, Massachusetts
02115, USA}

\author{Yang-Yu Liu}
\email{yyl@channing.harvard.edu}
\affiliation{Channing Division of Network Medicine, Brigham and
Women's Hospital and Harvard Medical School, Boston, Massachusetts
02115, USA}
\affiliation{Center for Cancer Systems Biology, Dana-Farber Cancer Institute, Boston, Massachusetts 02115, USA}%

\date{\today}

\begin{abstract}
Explaining biodiversity in nature is a fundamental problem in ecology.
An outstanding challenge is embodied in the so-called Competitive
Exclusion Principle: two species competing for one limiting resource
cannot coexist at constant population densities, or more generally,
the number of consumer species in steady coexistence cannot exceed
that of resources. The fact that competitive exclusion is rarely
observed in natural ecosystems has not been fully understood. Here we
show that by forming chasing triplets among the consumers and
resources in the consumption process, the Competitive Exclusion
Principle can be naturally violated. The modeling framework developed here is broadly applicable and can be used to explain the biodiversity of many consumer-resource
ecosystems and hence deepens our understanding of biodiversity in nature.
\end{abstract}

\maketitle


\section{\label{sec:level1}Introduction}

In Darwin's theory of evolution, survival of the fittest,
i.e., the less competitive species die out, implicates the notion of
competition exclusion~\cite{RN439}. In 1928, Volterra illustrated mathematically
that when two species compete for a single resource, one must die out unless the hunting to death rate ratio is exactly
the same for the two competing species
~\cite{RN440}. Those results were absorbed in the famous 
Competition Exclusion Principle (CEP)~\cite{RN441,RN459, RN440,RN448},
also named as Gause's Law~\cite{RN459}: two species
competing for one type of resource cannot coexist at steady state.
In the 1960s, 
MacArthur and 
Levins extended this
principle to ecosystems with arbitrary number of resource species
~\cite{RN445,RN446,RN444}.
Consider $M$ types of consumer
species competing for $N$ types of resources. Each consumer
can feed on one or multiple types of resources. Consumers do not
directly interact with each other via other mechanisms except
competing for the resources. According to the
CEP~\cite{RN445,RN446,RN444}, at steady state the number of coexisting
species of consumers cannot exceed that of resources, i.e., $M \le
N$ (see also Fig.S1).
%

\begin{figure*}[ht!]
\includegraphics[width=0.9\linewidth]{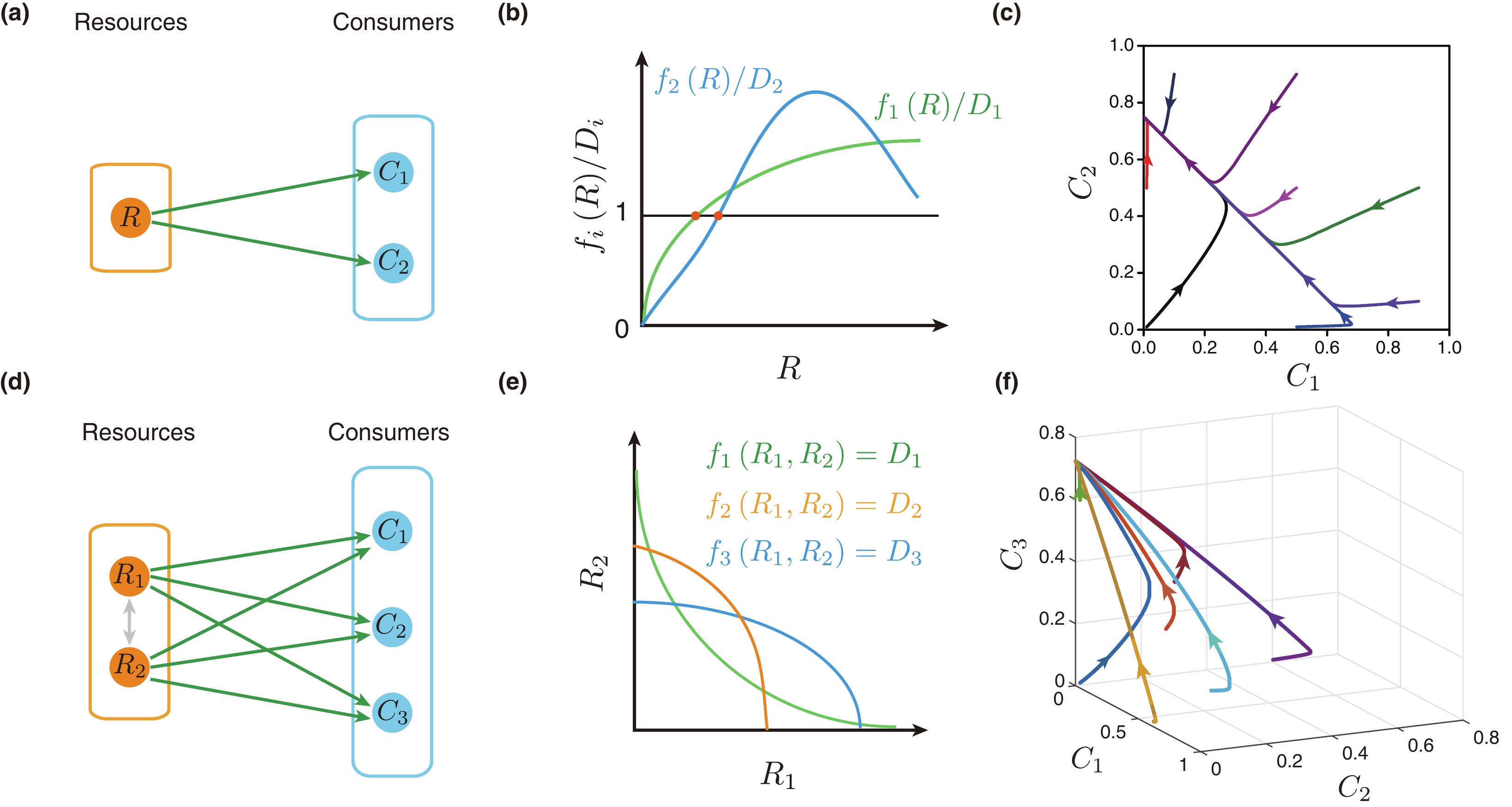}%
\caption{Classical proof of the competitive exclusion principle~\cite{RN445,RN446,RN444}.
(a) The scenario of two consumer species ($M=2$) and one resource
species ($N=1$). The green arrows denote the biomass flow among the consumption relationships.
(b) At steady state, if the two consumer species
coexist, then according to Eq.\ref{eq:1}, ${f_i}\left( R \right)/{D_i}
= 1$ ($i =1, 2$). This requires that the following three
lines $y = {f_i}\left( R \right)/{D_i} $ ($i =1, 2$) and $y = 1$ intersect
at a single point, which normally cannot happen.
(c) Representative trajectories of the two consumer species, which cannot coexist at
steady state when $N=1$.
Here ${f_i}\left( R \right) = \alpha{R}$ ($i$=1, 2); $g\left( {R,{C_1},{C_2}} \right) = R\left( {r^{\left(0\right)} -\beta^{\left( 0 \right)} R - \beta^{\left( 1 \right)}{C_1}- \beta^{\left( 2 \right)}{C_2}} \right)$; $\alpha=\beta^{\left( 0 \right)}=\beta^{\left( 1 \right)}=\beta^{\left( 1 \right)}=\beta^{\left( 2 \right)}=1$, ${D_1} = 0.0006$ and ${D_2} = 0.0005$. All
trajectories start at $R=0.01$.
(d) The scenario of three consumer species ($M=3$) and two resource
species ($N=2$). Predation or other interactions are forbidden among consumers but allowed (denoted by grey arrows) among resources. 
(e) If the three consumer species coexist at steady state, then
according to Eq.\ref{eq:2}, ${f_i}\left( {{R_1},{R_2}}
\right) = {D_i}$, ($i=1, 2, 3$). Generically, three curves would not
intersect at exactly the same point, hence the three consumer species
cannot coexist at steady state.
(f) Representative trajectories of the three consumer species, which
cannot all coexist at steady state (see Fig.S2 for the case that two
of the three consumer species coexist). 
Here ${f_i}\left( {{R_1},{R_2}} \right) = \alpha _i^{\left( 1 \right)}{R_1} + \alpha _i^{\left( 2 \right)}{R_2}$ ($i$ =1, 2, 3); ${g_j}\left(
{{R_1},{R_2},{C_1},{C_2},{C_3}} \right) = {R_j}\left( {r_j^{\left( 0
\right)} - \beta _j^{\left( 0 \right)} {R_j} - \beta _j^{\left( 1 \right)}{C_1} - \beta
_j^{\left( 2 \right)}{C_2} - \beta _j^{\left( 3 \right)}{C_3}}
\right)$ ($j$=1, 2); ${D_1} = 0.0006$, ${D_2} = 0.0005$, ${D_3} =
0.0004$, $\alpha _1^{\left( 1 \right)} = 0.0013$, $\alpha _2^{\left(
1 \right)} = 0.0011$, $\alpha _1^{\left( 2 \right)} = 0.001$,
$\alpha _2^{\left( 2 \right)} = 0.0009$, $r_1^{\left( 0 \right)} =
1.01$, $r_2^{\left( 0 \right)} = 1$, $ \beta _1^{\left( 0 \right)}=\beta _2^{\left( 0 \right)}=1$, $\beta _1^{\left( 1 \right)} =
1.3$, $\beta _2^{\left( 1 \right)} = 1$, $\beta _1^{\left( 2 \right)}
= 1.1$, $\beta _2^{\left( 2 \right)} = 0.9$, $\alpha _3^{\left( 1
\right)} = 0.0009$, $\alpha _3^{\left( 2 \right)} = 0.0013$, $\beta
_1^{\left( 3 \right)} = 0.9$ and $\beta _2^{\left( 3 \right)} = 1.3$. For
the initial condition, we set $R_1=0.01$ and $R_2=0.01$ for all
trajectories. 
}
\label{fig:1}
\end{figure*}

The classical proof~\cite{RN445,RN446,RN444} of the CEP is demonstrated
in Fig.\ref{fig:1}. Consider the simplest case: $M=2$ and $N=1$ (Fig.\ref{fig:1}a), i.e., two
consumer species $C_1$ and $C_2$ compete for one type of resource $R$.
The generic population dynamics of this consumer-resource ecosystem
can be described as follows:
\begin{equation}
\left\{ \begin{array}{l}
\dot{C_i} = {C_i}\left( {{f_i}\left( R \right) - {D_i}} \right), \quad
i=1,2; \\
\dot{R} = g\left( {R,{C_1},{C_2}} \right).
\end{array} \right.
\label{eq:1}
\end{equation}
Here $f_i$ and $g$ are unspecified functions, $D_i$ stands
for mortality rate of the consumer $C_i$. At steady state, if the two
consumer species coexist, we have ${f_i}\left( R \right) = D_i$,
$i=1,2$. This requires that the two curves
$y={f_1}\left( R \right)/{D_1}$ and $y={f_2}\left( R \right)/{D_2}$ should cross the line
$y=1$ at the same point, which is typically impossible
(Fig.\ref{fig:1}b), unless the model parameters satisfy certain
constraint (with Lebesgue measure zero, see Fig.S3). Hence the two
consumer species cannot coexist at steady state
(Fig.\ref{fig:1}c).
In the case of $M=3$ and $N=2$, the general
population dynamics Model Can be written as 
\begin{equation}
\left\{ \begin{array}{l}
\dot{C_i} = {C_i}\left( {{f_i}\left( {{R_1},{R_2}} \right) - {D_i}}
\right) , \quad
i=1,2,3; \\
\dot{R_j} = {g_j}\left( {{R_1},{R_2},{C_1},{C_2},{C_3}} \right), \quad j=1,2.
\end{array} \right.
\label{eq:2}
\end{equation}
Here $f_i$ and $g_j$ are unspecified functions, $D_i$ represents the
mortality rate of the consumer $C_i$. Similar proof strategy used in
the case of $M=2$ and $N=1$ can be applied here (see Fig.\ref{fig:1}d-f),
or more complicated cases with any positive $N$ and
$M$~\cite{RN445}.

Interestingly, 
an astonishing level of biodiversity has been witnessed in most natural ecosystems.
In aquatic biology, Hutchinson first proposed the paradox of the plankton: a limited number of resource types supports an unexpectedly large number of plankton species~\cite{RN442}. In tropical rainforests, one gram of soil contains a spectacular 2,000 to 18,000 distinct microbial genomes~\cite{RN601}. Explaining biodiversity is a great challenge in ecology.
In the past five decades, many mechanisms have been proposed to overcome the limitation on biodiversity set by CEP
~\cite{RN442,RN461,RN462,RN463,RN460,RN443,RN451,RN452,RN425,RN485,RN486,RN515,RN398,RN454,RN464,RN465,RN456,RN453}.
Some argued that ecosystem never approaches steady state due to
temporal~\cite{RN442,RN461,RN462} or spatial factors
~\cite{RN463,RN460}, or species self-organized
dynamics~\cite{RN443,RN451,RN452}.
Some considered special cases when the system parameters satisfy
certain constraints~\cite{RN440,RN425}. The rest considered
aspects such as cross-feeding~\cite{RN485,RN486,RN515}, toxin~\cite{RN398}, rock-paper-scissors relation~\cite{RN454,RN464,RN456}, complex
interactions~\cite{RN464,RN465,RN456} or co-evolution~\cite{RN453},
etc. (see SI Sec.II.A for details).
%
%
We emphasize that none of the existing mechanisms can generically explain the
violation of CEP at steady state~\cite{RN448,RN449}, 
without assuming any special model parameters.
Here we present a mechanism that considers the details of the consumption process. We find that
forming chasing triplets among the consumers and resources can
naturally break the CEP at steady state and hence facilitate biodiversity.

\section{\label{sec:level1}Modeling the consumption process}
\subsection{\label{sec:level2} Chasing-pair scenario}
\begin{figure}[ht!]
\includegraphics[width=0.8\linewidth]{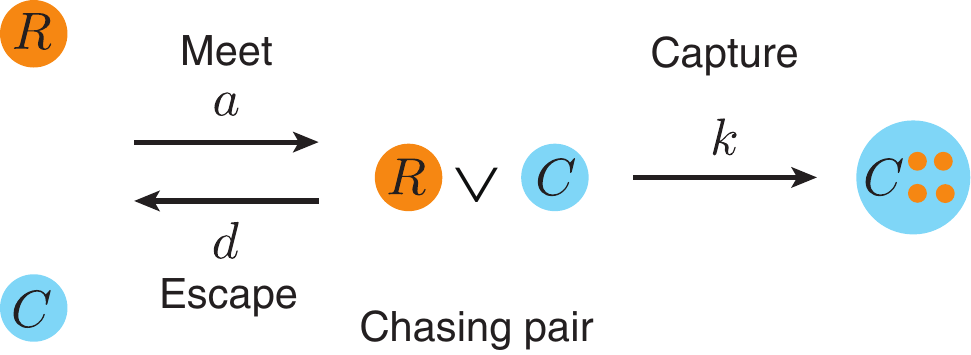}%
\caption{Schematic of the consumption process between consumers and resources. 
Here $a$ is the encounter rate between a consumer and a resource to form a chasing pair; 
$d$ is the escape rate of a resource out of a chasing pair; 
$k$ is the capture rate of consumers in a chasing pair.
}
\label{fig:2}
\end{figure}
Let's consider the consumption process between the consumers and
resources. The consumers are biotic, while the resources can be either biotic or abiotic (see Fig.\ref{fig:2}). 
We explicitly consider that the population structure of consumers and resources:
some are wandering around freely, some are chasing each other. When a
consumer meets a resource with rate $a$, they form a chasing pair,
denoted as ${R^{\left( {\rm{P}} \right)}} \vee {C^{\left( {\rm{P}}
\right)}}$, where the superscript `P' stands for `pair'. 
The resource can either ``escape" with rate $d$ or be caught and consumed by the consumer with rate $k$. For abiotic resources, the ``escape” rate corresponds to that the consumer fails to capture the resource in a chasing pair, which is analogous to a non-effective collision in chemical reactions.
Such a consumption kinetics commonly takes
the Michaelis-Menten form:
\begin{equation}
kC\frac{R}{{R + K}},
\label{eq:N1}
\end{equation}
with $K \equiv \frac{{d+ k}}{a}$, which corresponds to the Holling's type-II functional
response~\cite{RN483} in ecology and is widely adopted in consumer-resource models
~\cite{RN443,RN458}. This form, in fact, agrees with the growth rate
function in the classical proof~\cite{RN445,RN446,RN444}, where
$f\left( R \right) = k\frac{R}{{R + K}}$. Nevertheless, the
Michaelis-Menten kinetics is a good approximation only if the resource
population is much larger than the consumer population, i.e., $R \gg
C$ (see SI Sec.III for details). When this condition is not satisfied,
the growth rate function follows $f\left( {R,C} \right)$~\cite{RN418}
rather than $f\left( R \right)$. The $C$-dependency in the growth rate
function invalidates the classical proof~\cite{RN445,RN446,RN444},
implying a potential mechanism to break the CEP.

\begin{figure*}[ht!]
\includegraphics[width=0.9\linewidth]{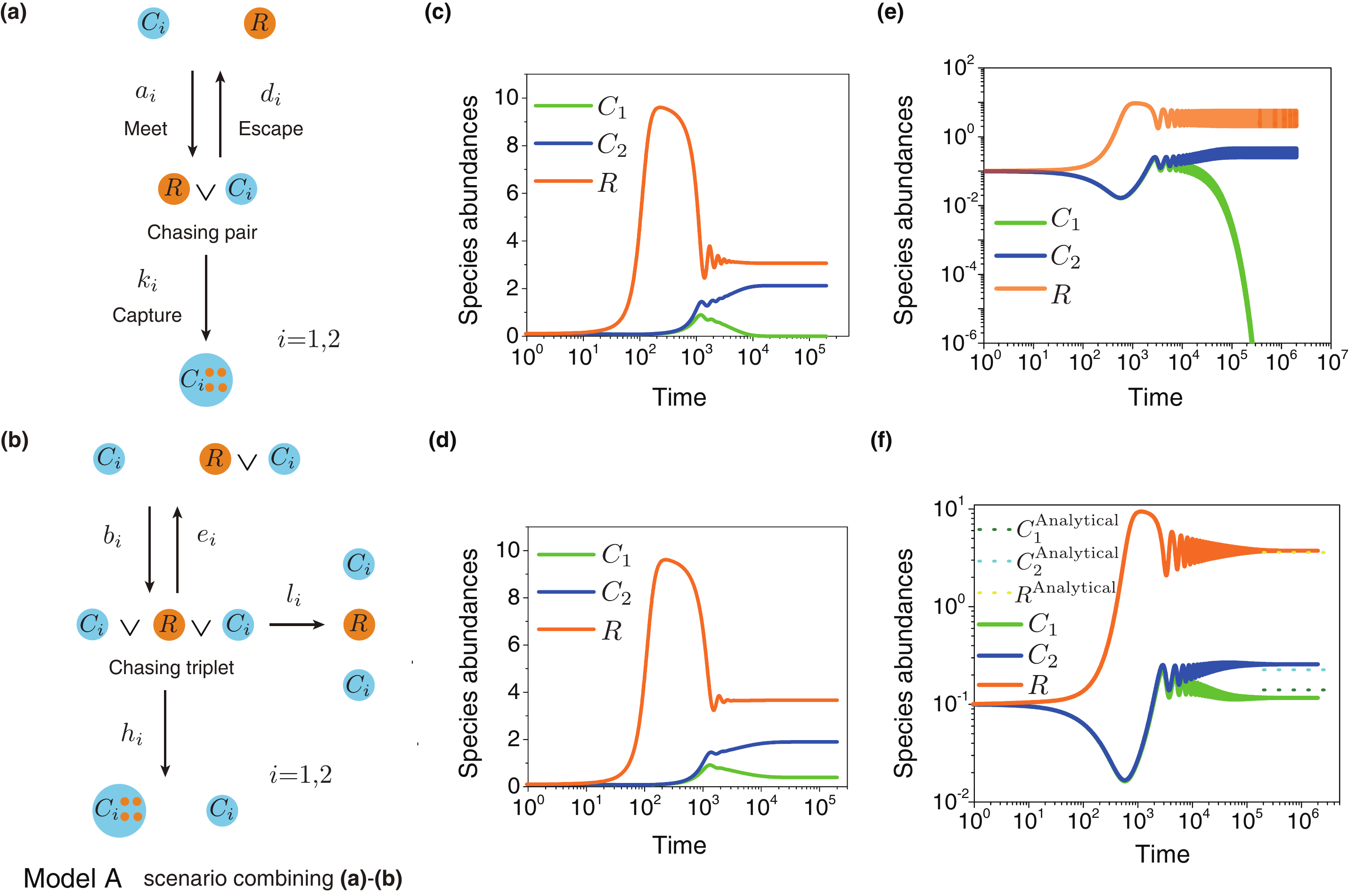}%
\caption{
Modeling the consumption process between consumers and resources
explicitly may naturally break the CEP.
For simplicity, we consider the case of two consumer species ($M=2$) and one biotic resource species ($N=1$, see Fig.S4 c, d for the case of abiotic resource species).
(a) Formation of a chasing pair between a consumer and a
resource. Here $a_i$ is the encounter rate between a consumer $C_i$ and a
resource to form a chasing pair $x_i$; $d_i$ is the escape rate of a
resource out of a chasing pair $x_i$; $k_i$ is the capture rate of
consumer $C_i$ in a chasing pair $x_i$.
(b) Formation of a chasing triplet among two consumers of the same species
and a resource. Here $b_i$ is the encounter rate between a consumer
species $C_i$ and an existing chasing pair $x_i$ to form a chasing
triplet $y_i$; $e_i$ and $l_i$ are the escape rates of a consumer
$C_i$ out of a chasing triplet $y_i$; $h_i$ is the capture rate of
consumer $C_i$ in a chasing triplet $y_i$. We denote the scenario combining chasing pair (a) and triplet (b) as Model A.
(c), (e) Time course of the abundances of two consumers species ($M=2$) and one resource species ($N=1$).
(c) In the presence of only chasing pairs, consumer species cannot coexist at steady state. (e) In the presence of only chasing pairs, only one type of consumer species exists for long, the oscillating dynamics resembles that of the classical predator-prey models ~\cite{RN526} . 
(d), (f) Time course of the abundances of two consumers species ($M=2$) and one
resource species ($N=1$). In the presence of chasing pairs and chasing triplet (Model A), consumer species can coexist at
steady abundances. The dotted lines in (f) are the steady-state analytical solutions (labeled with superscript `Analytical') calculated in Eqs.\ref{eq:10}-\ref{eq:12}.
(c) and (e) were simulated from Eq.\ref{eq:4}, where $g = R{R_0}(1 - R/{K_0}) - ({k_1}{x_1} + {k_2}{x_2})$.
(d) and (f) were simulated from Eqs.\ref{eq:5}-\ref{eq:6}. 
In (c)-(f): $D_2=0.005$, $K_0=10$, $a_i=0.1$, $d_i=0.1$, $k_i=0.1$ ($i$=1, 2); the initial abundances of $\left( {R,{C_1},{C_2}} \right)$ are $(0.1,0.1, 0.1)$.
In (c), (d):$R_0=0.05$, $D_1=1.08D_2$, $w_i=0.1$ ($i$=1, 2).
In (e), (f):$R_0=0.01$, $D_1=1.01D_2$, $w_i=0.08$ ($i$=1, 2) . 
In (d), (f):$b_i=0.1$, $e_i=0.1$, $h_i=0.1$, $l_i=0.1$ ($i$=1, 2).
}
\label{fig:3}
\end{figure*}
\subsection{\label{sec:level2}Ephemeral consumption process can influence the population dynamics}
\begin{figure*}[ht!]
\includegraphics[width=0.7\linewidth]{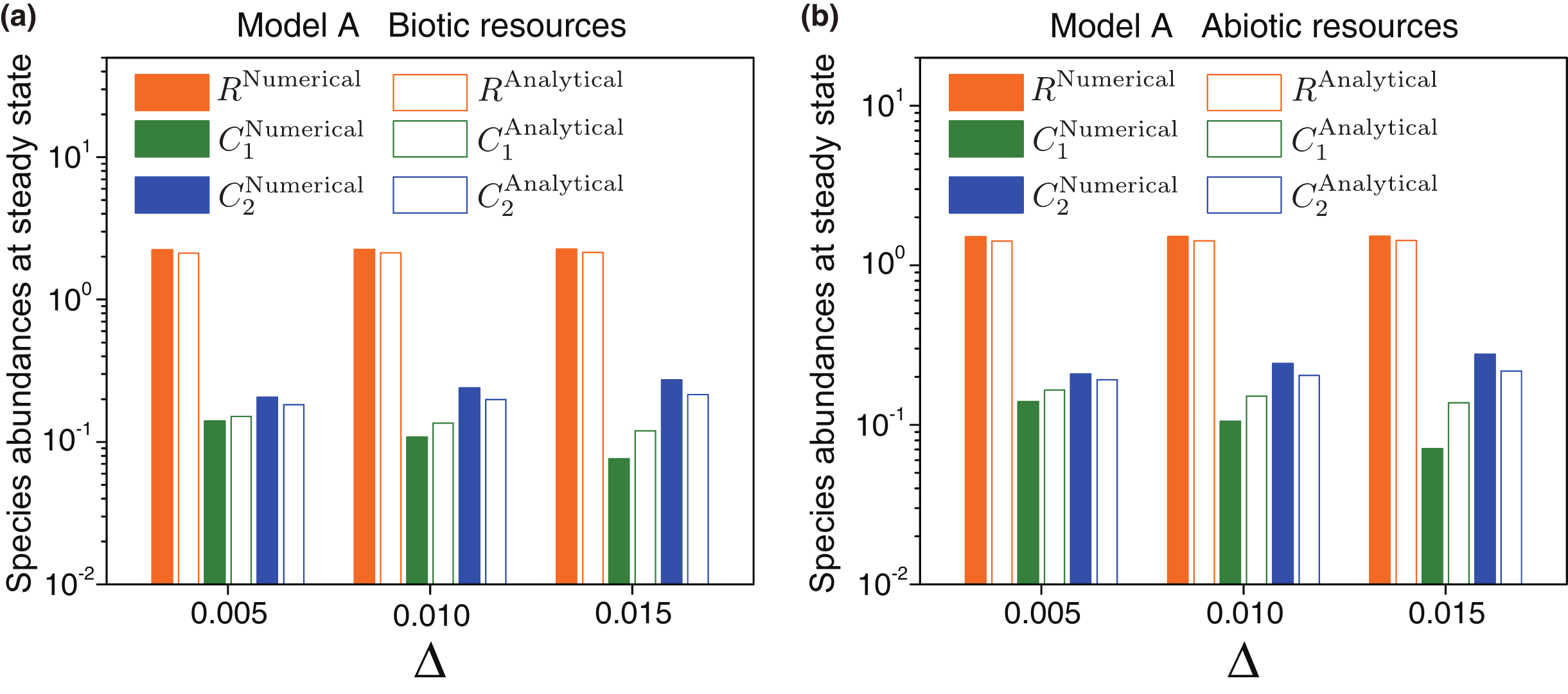}%
\caption{Comparison between analytical solutions and numerical results for the steady-state species abundances in Model A.
(a) biotic resources;
(b) abiotic resources.
Color bars are numerical results while hollow bars are analytical solutions. 
(a)-(b) the numerical results (labeled with superscript `Numerical') were calculated from Eqs.\ref{eq:5}-\ref{eq:6}, while the analytical results (labeled with superscript `Analytical') were calculated from Eqs.\ref{eq:10}-\ref{eq:13}.
In (a), (b): $\Delta \equiv ({D_1} - {D_2})/{D_2}$, $a_i=0.1$, $d_i=0.1$, $k_i=0.1$, $b_i=0.1$, $e_i=0.1$, $h_i=0.1$, $l_i=0.1$, $w_i=0.1$ ($i$=1, 2).
In (a): $D_2=0.005$, $K_0=10$, $R_0=0.01$.
In (b): $D_2=0.004$, $K_0=5$, $R_0=0.02$.
}
\label{fig:4}
\end{figure*}
For consumer species, the time-scale of consumption process is generally much faster than that of the birth and death processes. 
Then, how can the consumption process influence the population dynamics? To clarify this, we consider a simple scheme as follows.
A consumer individual of species $C$ was born with a mass of $m_C^{{\rm{new}}}$. When its mass increases to a critical value $m_C^{{\rm{birth}}}$, 
it would immediately give birth to a new individual with mass $m_C^{{\rm{new}}}$ and itself owns a mass of $m_C^{{\rm{mother}}}$ (the birth process). 
Due to the conservation of mass, $m_C^{{\rm{birth}}}=m_C^{{\rm{new}}}+m_C^{{\rm{mother}}}$. We use $D$ to denote the mortality rate of consumer species $C$ (the death process). Each time a consumer individual eats up a resource individual (from species $R$), it gains a incremental mass of ${m_\Delta }$.
Here we use Fig.2 to describe the consumption process. Denote the total mass of consumer species $C$ as $M_C$, 
then the population dynamics of $M_C$ follows:$\dot{M_C} = kx{m_\Delta } - D{M_C}$, where ${x \equiv R^{\left( {\rm{P}} \right)}} \vee {C^{\left( {\rm{P}}
\right)}}$ represents the chasing pair. 
Then, the population dynamics of the system can be described as follows:
\begin{equation}
\left\{ \begin{array}{l}
\dot{x} = {a}{R^{\left( {\rm{F}} \right)}}C^{\left( {\rm{F}}
\right)} - \left( {{d} + {k}} \right){x},\\
\dot{C} = {w}{k}{x} - {D}{C},\\
\dot{R} = g\left( {R,{x}, C} \right).
\end{array} \right.
\label{eq:3}
\end{equation}

Here, $ g\left( {R,{x}, C} \right)$ is a unspecified function. Consumers and resources that are freely wandering around are denoted as $C^{\left( {\rm{F}} \right)}$ and $R^{\left({\rm{F}} \right)}$, respectively, 
where the superscript `F' stands for `freely wandering'. $w$ is a biomass conversion ratio: the reciprocal of the number of resource individuals to be consumed to produce a new-born consumer. 
Generally speaking, $w$ is of the order of $\frac{{{m_\Delta }}}{{m_C^{{\rm{new}}}}} \ll 1$. From the population dynamics equation of $C$ (in Eq.\ref{eq:3}), although the kinetic parameters 
for consumption process is generically much larger than that of death process: $a,k,d \gg D$ (i.e., the consumption process is ephemeral), $x$ can be quite comparable to $C$ due to a small $w$, hence the consumption process can still influence the population dynamics, and should be explicitly considered in our modeling framework.

\subsection{\label{sec:level2} Forming Chasing-pair still cannot break the CEP}

Interestingly, we find that the presence of chasing pair and the
$C$-dependent growth rate functions are still not
enough to break the CEP. For example, in case $M=2$ and
$N=1$ (Fig.\ref{fig:3}a), the population dynamics of the system can be
described as follows: 
\begin{equation}
\left\{ \begin{array}{l}
\dot{x_i} = {a_i}{R^{\left( {\rm{F}} \right)}}C_i^{\left( {\rm{F}}
\right)} - \left( {{d_i} + {k_i}} \right){x_i},\\
\dot{C_i} = {w_i}{k_i}{x_i} - {D_i}{C_i},\\
\dot{R} = g\left( {R,{x_1},{x_2},{C_1},{C_2}} \right),
\end{array} \right.
\label{eq:4}
\end{equation}
with $i=1,2$.
Here consumers and resources that are freely wandering around are
denoted as $C_i^{\left( {\rm{F}} \right)}$ and $R^{\left(
{\rm{F}} \right)}$, respectively. 
The variable ${x_i} \equiv {R^{\left( {\rm{P}} \right)}} \vee
C_i^{\left( {\rm{P}} \right)}$ represents the chasing pair, $a_i$ is
the encounter rate between a consumer $C_i$ and a resource to form a chasing
pair $x_i$, $d_i$ is the ``escape" rate of a resource out of a chasing
pair $x_i$, and $k_i$ is the capture rate of consumer $C_i$ in a chasing pair $x_i$. If the
two consumers can coexist, we prove that the steady-state
equations yield ${f_i}\left( {{R^{\left( {\rm{F}} \right)}}}
\right)/{D_i} = 1$, which corresponds to parallel planes in
the $\left( {{C_1},{C_2},{R^{\left( {\rm{F}} \right)}}} \right)$
coordinate system (Fig. S6b), rendering coexistence impossible (Fig.\ref{fig:3}c, e and
Fig S6, see SI Sec.IV-V for details).

\subsection{\label{sec:level2} Consumption process with chasing-triplet}

Pack-hunting is prevalent across different organisms in the wild ~\cite{RN519,RN520,RN521,RN522,RN523,RN524,RN525}, which means that two or more consumer individuals can chase the same resource individual simultaneously~\cite{RN519,RN520,RN521,RN522,RN523,RN524,RN525}. To take this into account, we revisit the consumption process and naturally extend the idea of chasing pair to chasing triplet, i.e., two consumers (within the same or from different species) can chase the same resource (Fig.\ref{fig:3}b,
\ref{fig:6}a and \ref{fig:6}b). For example, in case $M=2$ and $N=1$, a consumer
$C_i$ can join an existing chasing pair 
\begin{equation}
{x_i} \equiv
{R^{\left( {\rm{P}} \right)}} \vee C_i^{\left( {\rm{P}} \right)}
\label{eq:N2}
\end{equation}
to form a chasing triplet 
\begin{equation}
{y_i} \equiv C_i^{\left( {\rm{T}} \right)} \vee
{R^{\left( {\rm{T}} \right)}} \vee C_i^{\left( {\rm{T}} \right)}
\label{eq:N3}
\end{equation}
(Fig.\ref{fig:3}b, denoted as Model A), where the superscript `T' stands for `triplet'. Those consumers and resources that are freely wandering around are still
denoted as $C_i^{\left( {\rm{F}} \right)}$ and $R^{\left(
{\rm{F}} \right)}$, respectively. The population of consumers and
resources are given by ${C_i} = C_i^{\left(
{\rm{F}} \right)} + {x_i} + 2{y_i}$ ($i=1, 2$) and $R = {R^{\left(
{\rm{F}} \right)}} + \sum\limits_{i = 1}^2 {\left( {{x_i} +
{y_i}} \right)}$, respectively. The population dynamics of the
system can be described as follows 
\begin{equation}
\left\{ \begin{array}{l}
\dot{x_i} = {a_i}{R^{\left( {\rm{F}} \right)}}C_i^{\left( {\rm{F}} \right)} - \left( {{d_i} + {k_i}} \right){x_i} - {b_i}{x_i}C_i^{\left( {\rm{F}} \right)} + {e_i}{y_i},\\
\dot{y_i} = {b_i}{x_i}C_i^{\left( {\rm{F}} \right)} - \left( {{h_i} + {e_i} + {l_i}} \right){y_i},\\
\dot{C_i} = {w_i}({k_i}{x_i} + {h_i}{y_i}) - {D_i}{C_i},\\
\dot{R} = g\left( {R,{x_1},{x_2},{y_1},{y_2},{C_1},{C_2}} \right),
\end{array} \right.
\label{eq:5}
\end{equation}
with $i=1,2$. Here $b_i$ is the encounter rate between a consumer
$C_i$ and an existing chasing pair $x_i$ to form a chasing triplet $y_i$; $e_i$ and $l_i$ are the
escape rates of a consumer $C_i$ out of a chasing triplet $y_i$ (Fig.\ref{fig:3}b); $h_i$ is
the capture rate of consumer $C_i$ in a chasing triplet $y_i$; $w_i$ and $D_i$ are the biomass conversion ratio and mortality rate of consumer $C_i$, respectively.

\subsection{\label{sec:level2} Forming Chasing-triplet can break the CEP}
In Eq.\ref{eq:5}, the explicit form of function $g\left( {R,{x_1},{x_2},{y_1},{y_2},{C_1},{C_2}} \right)$ has not been specified. Here we assume that the dynamics of the resources follow the same construction principle as that in the classical MacArthur's consumer-resource model~\cite{RN467,RN468}.
Then,
\begin{widetext}
\begin{equation}
g\left( {R,{x_1},{x_2},{y_1},{y_2},{C_1},{C_2}} \right)= 
\begin{cases}
R{R_0}(1 - R/{K_0}) - ({k_1}{x_1} + {k_2}{x_2}) - ({h_1}{y_1} + {h_2}{y_2}), & \text{for biotic resources.} \\
{R_0}(1 - R/{K_0}) -({k_1}{x_1}+ {k_2}{x_2}) - ({h_1}{y_1} + {h_2}{y_2}), & \text{for abiotic resources.}
\end{cases}
\label{eq:6}
\end{equation}
\end{widetext}

Using dimensional analysis, we make all parameters dimensionless (see SI Sec VII for details). For convenience, below we still use the same parameter notations, yet they are all dimensionless. Actually we can reduce two parameters in both the biotic and abiotic resource cases: ${K_0}$ and ${D_2}$. These two parameters can be set as either 1 or any other arbitrary positive real numbers (see SI Sec VII for details). In our calculations, we set ${K_0=10}$ and ${D_2=0.005}$ for biotic resource cases while ${K_0=5}$ and ${D_2=0.004}$ for abiotic resource cases. In the numerical simulations of Model A (Fig.\ref{fig:3}a, b), we find that two consumer species can achieve steady coexistence when there is only one type of resource (Fig.\ref{fig:3}d, f, Fig.S4d). 
\begin{figure*}[ht!]
\includegraphics[width=0.75\linewidth]{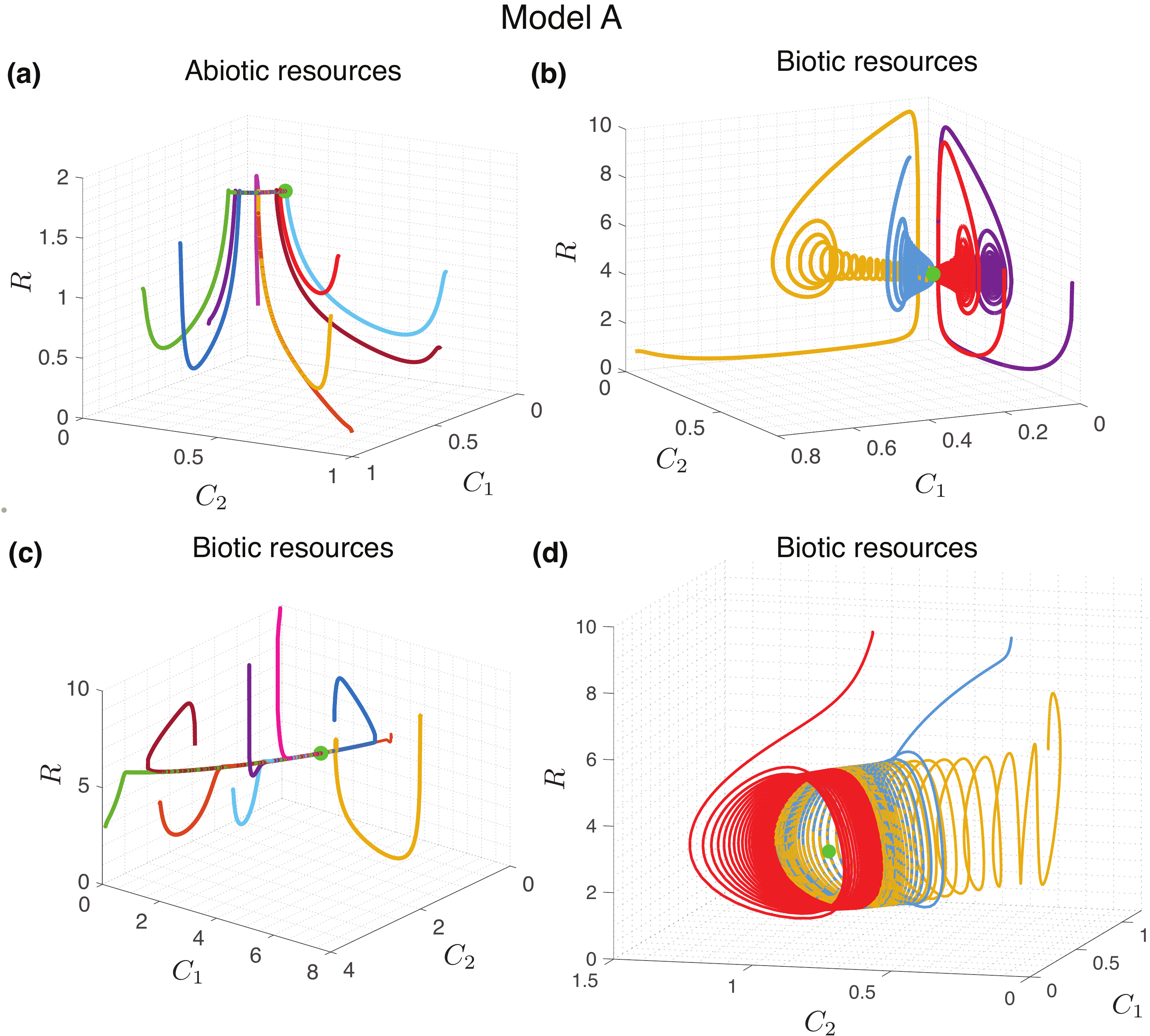}%
\caption{Different types of coexistence trajectories in the state space of model-A.
(a) Abiotic resource case, the coexistence state (green dot) is globally attracting.
(b)-(d) Biotic resource cases, green dot marks the fixed point. (b), (c) The coexistence state is globally attracting. (d) The coexistence state is unstable; all trajectories attract to a stable limit cycle. 
%
(a)-(d) were simulated from Eqs.\ref{eq:5}-\ref{eq:6}. 
In (a)-(d): $a_i=0.1$, $d_i=0.1$, $k_i=0.1$, $b_i=0.1$, $e_i=0.1$, $h_i=0.1$, $l_i=0.1$, $w_i=0.1$ ($i$=1, 2).
In (a), (c), (d): $w_i=0.1$ ($i$=1, 2).
In (a): $R_0=0.01$, $K_0=5$, $D_2=0.004$, $D_1=1.01D_2$.
In (b)-(d): $K_0=10$, $D_2=0.005$.
In (b): $R_0=0.01$, $D_1=1.01D_2$, $w_i=0.08$ ($i$=1, 2).
In (c): $R_0=0.08$, $D_1=1.08D_2$.
In (d): $R_0=0.03$, $D_1=1.03D_2$.
}
\label{fig:5}
\end{figure*}

To figure out how can these species steadily coexist, we resort to the steady state solution in Eq.\ref{eq:5}.
Define
\begin{equation}
\left\{ \begin{array}{l}
P_1^{\left( i \right)} = \left( {2{d_i} + 2{k_i} - {h_i} - {l_i}} \right){b_i},\\
P_2^{\left( i \right)} = \left( {{h_i} + {e_i} + {l_i}} \right){a_i},\\
P_3^{\left( i \right)} = {{\left( {{d_i} + {k_i}} \right)} \mathord{\left/
{\vphantom {{\left( {{d_i} + {k_i}} \right)} {{a_i}}}} \right.
\kern-\nulldelimiterspace} {{a_i}}},\\
P_4^{\left( i \right)} = \left( {{h_i} + {l_i}} \right){b_i},\\
P_5^{\left( i \right)} = {{\left( {{h_i} + {l_i}} \right)} \mathord{\left/
{\vphantom {{\left( {{h_i} + {l_i}} \right)} {{a_i}}}} \right.
\kern-\nulldelimiterspace} {{a_i}}},
\end{array} \right.\quad i=1, 2.
\label{eq:8}
\end{equation}
Note that $0 \le {x_i},{y_i} \le \min \left( {{C_i},R} \right)$, 
then
\begin{widetext}
\begin{equation}
\left\{ \begin{array}{l}
{x_i} = \frac{\sqrt {{{[P_2^{\left( i \right)}(P_3^{\left( i \right)} + {R^{\left( {\rm{F}} \right)}}) + P_4^{\left( i \right)}{C_i}]}^2} + 4P_1^{\left( i \right)}P_2^{\left( i \right)}{R^{\left( {\rm{F}} \right)}}{C_i}}
- [P_2^{\left( i \right)}(P_3^{\left( i \right)} + {R^{\left( {\rm{F}} \right)}}) + P_4^{\left( i \right)}{C_i}]
}{{2P_1^{\left( i \right)}}},\\
{y_i} = \frac{{{R^{\left( {\rm{F}} \right)}}{C_i} - (P_3^{\left( i \right)} + {R^{\left( {\rm{F}} \right)}}){x_i}}}{{2{R^{\left( {\rm{F}} \right)}} + P_5^{\left( i \right)}}},\quad i=1, 2.
\end{array} \right.
\label{eq:9}
\end{equation}
\end{widetext}
For arbitrary parameters, there is no analytical solution. However, when the abundance of resources are much larger than that of consumers, i.e., $R \gg {C_1},{C_2}$, which applies to almost all cases in the wild, then $R \approx {R^{\left( {\rm{F}} \right)}}$. Combining these results with $\dot{C_i}=0$ ($i$=1, 2) and $\dot{g}=0$, then we have 
\begin{equation}
{C_1} = \frac{{P_2^{\left( 1 \right)}R\left( {{{{w_1}{k_1}} \mathord{\left/
{\vphantom {{{w_1}{k_1}} {{D_1}}}} \right.
\kern-\nulldelimiterspace} {{D_1}}} - 1} \right) - P_2^{\left( 1 \right)}P_3^{\left( 1 \right)}}}{{P_4^{\left( 1 \right)}}},
\label{eq:10}
\end{equation}
\begin{figure*}[ht!]
\includegraphics[width=0.99\linewidth]{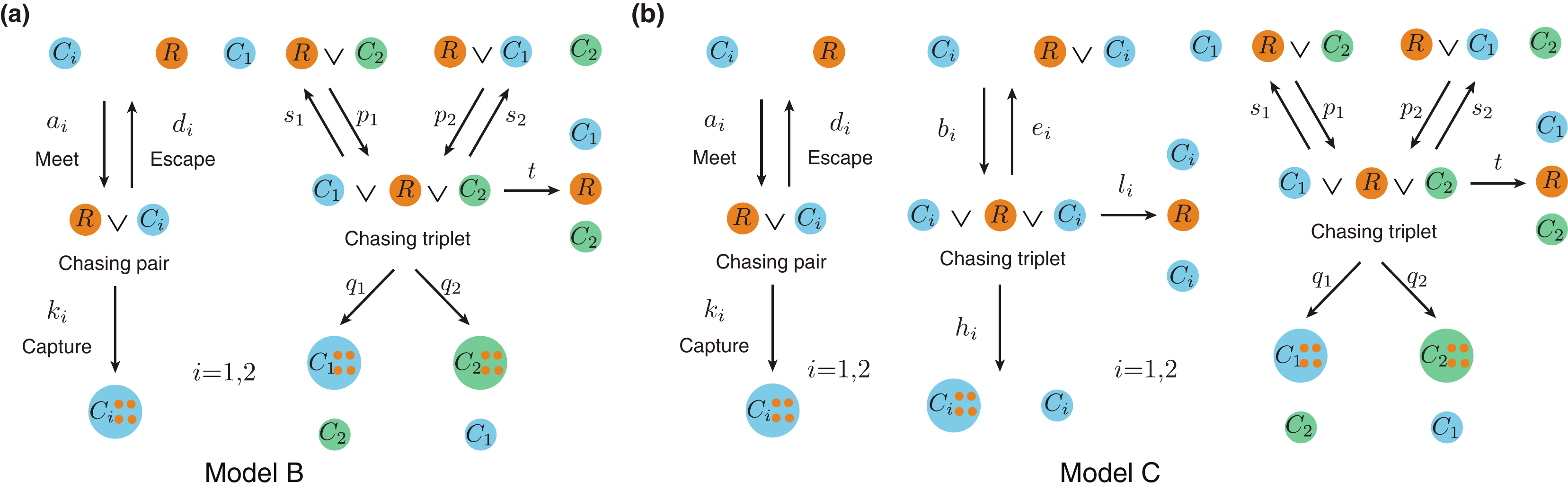}%
\caption{Schematic of the consumption process between consumers and resources.
(a) Model B, scenarios involving both chasing pair and triplet, a triplet consists of two consumers of the same species and a resource (denote as hetero chasing triplet). (b) Model C, scenarios involving both chasing pair and triplet, a triplet consists of two consumers of the same (denote as homo chasing triplet) or different species and a resource. In (a)-(b): $a_i$ is the encounter rate between a consumer and a resource to form a chasing pair, $d_i$ is the escape rate of a resource out of a chasing pair, $k_i$ is the capture rate of consumers in a chasing pair; while $p_i$ is the encounter rate between a consumer and an existing chasing pair to form a hetero chasing triplet, $s_i$ and $t$ are the escape rates of a consumer out of a hetero chasing triplet, $q_i$ is the capture rate of consumers in a hetero chasing triplet. In (b): $b_i$ is the encounter rate in forming a homo chasing triplet; $e_i$ and $l_i$ are the escape rates of a consumer out of a homo chasing triplet; $h_i$ is the capture rate of consumer in a homo chasing triplet. 
}
\label{fig:6}
\end{figure*}

\begin{equation}
{C_2} = \frac{{P_2^{\left( 2 \right)}R\left( {{{{w_2}{k_2}} \mathord{\left/
{\vphantom {{{w_2}{k_2}} {{D_2}}}} \right.
\kern-\nulldelimiterspace} {{D_2}}} - 1} \right) - P_2^{\left( 2 \right)}P_3^{\left( 2 \right)}}}{{P_4^{\left( 2 \right)}}}.
\label{eq:11}
\end{equation}
For biotic resources,
\begin{equation}
R = \frac{{\sqrt {{P_6}^2 + 4{P_6}{P_7}} - {P_6}}}{2},
\label{eq:12}
\end{equation}
where ${P_6} \equiv \frac{{{K_0}}}{{{r_0}}}[\frac{{P_2^{\left( 1 \right)}({k_1} - {{{D_1}} \mathord{\left/
{\vphantom {{{D_1}} {{w_1}}}} \right.
\kern-\nulldelimiterspace} {{w_1}}})}}{{P_4^{\left( 1 \right)}}} + \frac{{P_2^{\left( 2 \right)}({k_2} - {{{D_2}} \mathord{\left/
{\vphantom {{{D_2}} {{w_2}}}} \right.
\kern-\nulldelimiterspace} {{w_2}}})}}{{P_4^{\left( 2 \right)}}} - {r_0}]$ and
${P_7} = \frac{{{K_0}}}{{{r_0}}}(\frac{{{D_1}}}{{{w_1}}}\frac{{P_2^{\left( 1 \right)}P_3^{\left( 1 \right)}}}{{P_4^{\left( 1 \right)}}} + \frac{{{D_2}}}{{{w_2}}}\frac{{P_2^{\left( 2 \right)}P_3^{\left( 2 \right)}}}{{P_4^{\left( 2 \right)}}})$ .
For abiotic resources, 
\begin{equation}
R = \frac{{{r_0} + (\frac{{{D_1}}}{{{w_1}}}\frac{{P_2^{\left( 1 \right)}P_3^{\left( 1 \right)}}}{{P_4^{\left( 1 \right)}}} + \frac{{{D_2}}}{{{w_2}}}\frac{{P_2^{\left( 2 \right)}P_3^{\left( 2 \right)}}}{{P_4^{\left( 2 \right)}}})}}{{\frac{{P_2^{\left( 1 \right)}(k - {D_1}/{w_1})}}{{P_4^{\left( 1 \right)}}} + \frac{{P_2^{\left( 2 \right)}(k - {D_2}/{w_2})}}{{P_4^{\left( 2 \right)}}} + \frac{{{r_0}}}{{{K_0}}}}}.
\label{eq:13}
\end{equation}

These results (Eqs.\ref{eq:10}-\ref{eq:13}) are the analytical solutions to the steady-state species abundances under the approximation that $R \gg {C_1},{C_2}$. A necessary condition for species coexistence is that $\min \left( {R,{C_1},{C_2}} \right) > 0$. In Fig.\ref{fig:3}f, we show the analytical solutions of biotic resource case, which agree well with the simulation results. 
In Fig.\ref{fig:4}, we compared the analytical solutions (Eqs.\ref{eq:10}-\ref{eq:13}, the approximate solutions) of both the biotic and abiotic resource case to the numerical results (the exact solutions) at steady state, which overall shows good consistency for both cases. Here we assign $D_i$ ($i$=1, 2) to be the only different parameter between consumer species $C_1$ and $C_2$, then $ \Delta \equiv ({D_1} - {D_2})/{D_2}$, the relative difference in mortality rate, measures the competitive differences between the two consumer species. In Fig.\ref{fig:4}, we find that the analytical solution is closer to the exact solution when two consumer species are similarly competitive, while it deviates more when the competitive difference between the two consumer species gets larger within the coexistence region (the analytical solution involves approximation, see SI Sec.V.B for details). Overall, the analytical solutions (Eqs.\ref{eq:10}-\ref{eq:13}) are good approximations to predict species abundances at steady state, while exact solutions are required to identify the boundary of parameter space for species coexistence.

Interestingly, there are several types of coexistence trajectories in phase space within the scenario of Model A, which involves chasing pair, and triplet formed between consumers of the same species. When the resource is abiotic, there is only one type of behavior: the coexistence state is globally attracting as long as the initial abundances of these species are non-zero, as shown in Fig.\ref{fig:5}a. However, in the case that the resource is biotic, the coexistence state can be either globally attracting (Fig.\ref{fig:5}b, c) or unstable, leading to a limit cycle (Fig.\ref{fig:5}d) (see Fig.S4b for the oscillating coexistence in time series). In some cases, the oscillations damps, and ends in the globally attracting fixed point, as shown in Fig.\ref{fig:5}b.

\subsection{Scenarios with other chasing triplet forms}
To fully take into account scenarios involving different forms of chasing triplet (with the presence of chasing pair), we further consider cases where the chasing triplet is formed between different species of consumers (denoted as Model B, see Fig.\ref{fig:6}a) or either between the same or different species (denoted as Model C, see Fig.\ref{fig:6}b). 

\begin{figure*}[ht!]
\includegraphics[width=0.8\linewidth]{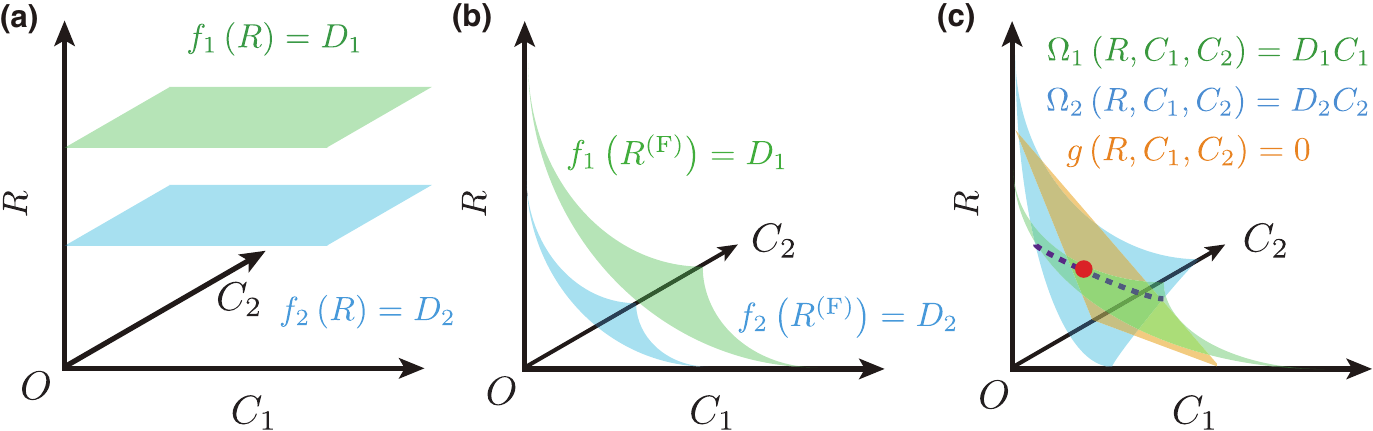}%
\caption{Intuitive explanation of why the formation of chasing can break the CEP. For simplicity, we consider the case of $M=2$ and $N=1$. 
(a) In the classical proof, the green plane and blue plane are parallel to each other, and thus do not have a common point. 
(b) In the model involving chasing pairs, the green surface and blue surface are still parallel to each other, and thus still do not have a common point (see Fig. S6b, SI Sec.IV-V for details). 
(c) In the model involving both chasing pairs and chasing triplets, the yellow, green and blue surfaces are not parallel to each other, and thus the green and the blue ones can have an intersection curve (shown in dashed purple), while the three surfaces can intersect at one point (shown in red) and thus facilitate coexistence.}
\label{fig:7}
\end{figure*}
In Model B (Fig.\ref{fig:6}a), the population dynamics can be written as follows:
\begin{equation}
\left\{ \begin{array}{l}
\dot{x_1} = {a_1}{R^{\left( {\rm{F}} \right)}}C_1^{\left( {\rm{F}} \right)} - \left( {{d_1} + {k_1}} \right){x_1} - {p_2}{x_1}C_2^{\left( {\rm{F}} \right)} + {s_2}z,\\
\dot {x_2} = {a_2}{R^{\left( {\rm{F}} \right)}}C_2^{\left( {\rm{F}} \right)} - \left( {{d_2} + {k_2}} \right){x_2} - {p_1}{x_2}C_1^{\left( {\rm{F}} \right)} + {s_1}z,\\
\dot {z} = {p_1}{x_2}C_1^{\left( {\rm{F}} \right)} + {p_2}{x_1}C_2^{\left( {\rm{F}} \right)} - \left( {{q_1} + {q_2} + {s_1} + {s_2}{\rm{ + }}t} \right)z,\\
\dot {C_i} = {w_i}({k_i}{x_i} + {q_i}z) - {D_i}{C_i}, \quad i=1,2;\\
\dot {R} = g\left( {R,x_1,x_2,z,{C_1},{C_2}} \right), 
\end{array} \right.
\label{eq:14}
\end{equation}
\begin{figure*}[ht!]
\includegraphics[width=0.99\linewidth]{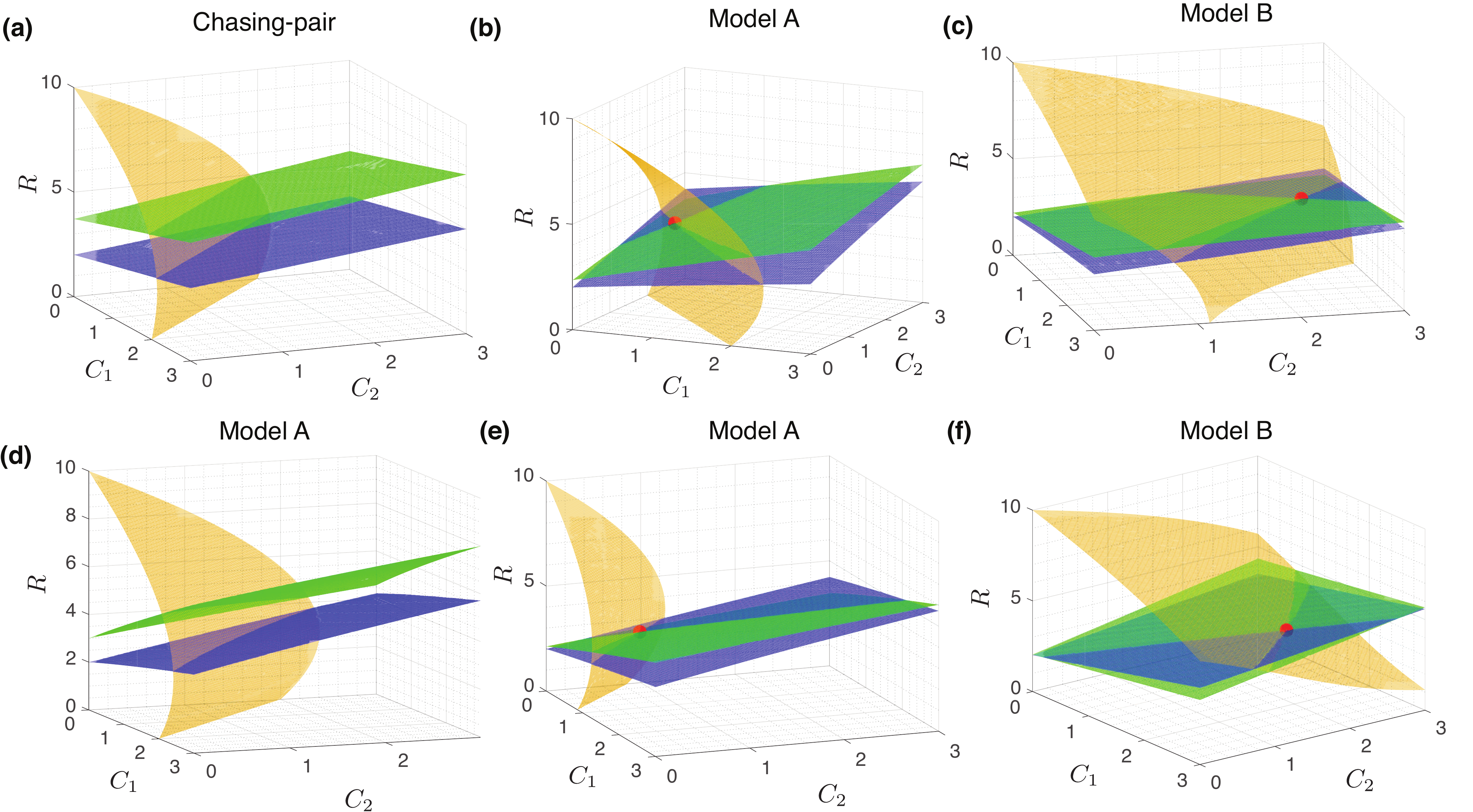}%
\caption{Demonstration of the intuitive explanation with numerical solutions. 
(a) In the scenario involving only chasing pair, numerical solutions confirm that the green surface and blue surface are parallel to each other.
(b), (c) In scenarios involving both chasing pair and triplet, numerical solutions confirm that the yellow, green and blue surfaces are not parallel to each other and definitely can have a common point (marked with red dot). 
(b) Numerical solutions of Model A (see Fig.\ref{fig:3} d for time series). 
(c) Numerical solutions of Model B (see Fig.S5a for time series). 
(d)-(f) Forming chasing triplet is not sufficient for species steady coexistence.
(d) Although the green surface and blue surface are not parallel to each other, yet they do not have an intersection curve in the first quadrant (i.e., $R,{C_1},{C_2} > 0$, see Fig.S8a for time series). 
(e) The intersection point (red dot) of the three surfaces is unstable. All trajectories end in a limit cycle (Fig.\ref{fig:5}d, see Fig.S4b for time series).
(f) The intersection point (red dot) of the three surface is unstable (see Fig. S8b for time series).
(a) was calculated from Eq.\ref{eq:4}, where $g = R{R_0}(1 - R/{K_0}) - ({k_1}{x_1} + {k_2}{x_2})$; (b), (d) and (e) were calculated from Eqs.\ref{eq:5}-\ref{eq:6}; (c) and (f) were calculated from Eqs.\ref{eq:14}-\ref{eq:15}. 
In (a)-(f): $a_i=0.1$, $d_i=0.1$, $k_i=0.1$, $w_i=0.1$ ($i$=1, 2); $D_2=0.005$, $K_0=10$.
In (a): $R_0=0.05$, $D_1=1.3D_2$.
In (b), (d), (e): $b_i=0.1$, $e_i=0.1$, $h_i=0.1$, $l_i=0.1$ ($i$=1, 2).
In (b): $R_0=0.05$, $D_1=1.08D_2$. 
In (d): $R_0=0.05$, $D_1=1.2D_2$. 
In (e): $R_0=0.03$, $D_1=1.03D_2$. 
In (c) and (f) $p_i=0.1$, $s_i=0.1$ ($i$=1, 2); $t=0.1$, $R_0=0.08$.
In (c): $D_1=1.05D_2$, $q_i=0.1$ ($i$=1, 2).
In (f): $D_1=1.01D_2$, $q_i=0.05$ ($i$=1, 2).
}
\label{fig:8}
\end{figure*}
where $x_i \equiv {R^{\left( {\rm{P}} \right)}} \vee C_i^{\left( {\rm{P}} \right)}$ represents the chasing pair, $z \equiv C_1^{\left( {\rm{T}} \right)} \vee {R^{\left( {\rm{T}} \right)}} \vee C_2^{\left( {\rm{T}} \right)}$ represents the chasing triplet, $C_i^{\left( {\rm{F}} \right)}$ ($i$=1, 2) and $R^{\left( {\rm{F}} \right)}$ stand for freely wandering consumers and resources, respectively. $a_i$, $d_i$, $k_i$, $p_i$, $q_i$, $s_i$ and $t$ stand for consuming process relevant parameters specified in the figure captions of Fig.\ref{fig:6}a. ${C_i} = C_i^{\left( {\rm{F}} \right)} + {x_i} + z$ ($i$=1, 2) and $R = {R^{\left( {\rm{F}} \right)}} + {x_1} + {x_2} + z$ are the populations of consumers and resources, respectively. $D_i$ represents the mortality rate of the consumer species, $w_i$ is the biomass conversion ratio. 
Assuming that the dynamics of resources $g$ follows the construction principle as that of the MacArthur's Model~\cite{RN467,RN468}, we have 
\begin{widetext}
\begin{equation}
g\left( {R,x_1,x_2,z,{C_1},{C_2}} \right) =
\begin{cases}
R{R_0}(1 - R/{K_0}) - ({k_1}{x_1} + {k_2}{x_2}) - ({q_1} + {q_2})z, & \text{for biotic resources.} \\
{R_0}(1 - R/{K_0})- ({k_1}{x_1} + {k_2}{x_2}) - ({q_1} + {q_2})z, & \text{for abiotic resources.}
\end{cases}
\label{eq:15}
\end{equation}
In Model C (Fig.\ref{fig:6}b), the population dynamics can be written as follows:
\begin{equation}
\left\{ \begin{array}{l}
\dot{x_1} = {a_1}{R^{\left( {\rm{F}} \right)}}C_1^{\left( {\rm{F}} \right)} - \left( {{d_1} + {k_1}} \right){x_1} - {b_1}{x_1}C_1^{\left( {\rm{F}} \right)}+ {e_1}{y_1} - {p_2}{x_1}C_2^{\left( {\rm{F}} \right)} + {s_2}z,\\
\dot {x_2} = {a_2}{R^{\left( {\rm{F}} \right)}}C_2^{\left( {\rm{F}} \right)} - \left( {{d_2} + {k_2}} \right){x_2} - {b_2}{x_2}C_2^{\left( {\rm{F}} \right)}+ {e_2}{y_2} - {p_1}{x_2}C_1^{\left( {\rm{F}} \right)} + {s_1}z,\\
\dot {y_i} = {b_i}{x_i}C_i^{\left( {\rm{F}} \right)} - \left( {{h_i} + {e_i} + {l_i}} \right){y_i},\quad i=1,2,\\
\dot {z} = {p_1}{x_2}C_1^{\left( {\rm{F}} \right)} + {p_2}{x_1}C_2^{\left( {\rm{F}} \right)} - \left( {{q_1} + {q_2} + {s_1} + {s_2}{\rm{ + }}t} \right)z,\\
\dot {C_i} = {w_i}({k_i}{x_i} + {h_i}{y_i} + {q_i}z )- {D_i}{C_i} ,\quad i=1,2,\\
\dot {R} = g\left( {R,x_1,x_2,y_1,y_2,z,{C_1},{C_2}} \right). 
\end{array} \right.
\label{eq:17}
\end{equation}
\end{widetext}
where $x_i \equiv {R^{\left( {\rm{P}} \right)}} \vee C_i^{\left( {\rm{P}} \right)}$ represents the chasing pair, $y_i \equiv C_i^{\left( {\rm{T}} \right)} \vee {R^{\left( {\rm{T}} \right)}} \vee C_i^{\left( {\rm{T}} \right)}$ and $z \equiv C_1^{\left( {\rm{T}} \right)} \vee {R^{\left( {\rm{T}} \right)}} \vee C_2^{\left( {\rm{T}} \right)}$ represent the chasing triplet, $C_i^{\left( {\rm{F}} \right)}$ ($i$=1, 2) and $R^{\left( {\rm{F}} \right)}$ stand for freely wandering consumers and resources, respectively. $a_i$, $b_i$, $d_i$, $e_i$, $h_i$, $k_i$, $l_i$, $p_i$, $q_i$, $s_i$ and $t$ stand for consuming process relevant parameters specified in the figure captions of Fig.\ref{fig:6}b. ${C_i} = C_i^{\left( {\rm{F}} \right)} + {x_i} + 2{y_i} + z$ ($i$=1, 2) and $R = {R^{\left( {\rm{F}} \right)}} + \sum\limits_{i = 1}^2 {\left( {{x_i} + {y_i}} \right)} + z$ are the populations of consumers and resources, respectively. 
Assuming that $g$ follows the construction principle as that of the MacArthur's Model~\cite{RN467,RN468}, we have 
\begin{widetext}
\begin{equation}
\begin{array}{l}
g\left( {R,x_1,x_2,y_1,y_2,z,{C_1},{C_2}} \right)= 
\begin{cases}
R{R_0}(1 - R/{K_0}) - ({k_1}{x_1} + {h_1}{y_1}) - ({k_2}{x_2} + {h_2}{y_2})- ({q_1} + {q_2})z, & \text{for biotic resources.} \\
{R_0}(1 - R/{K_0}) - ({k_1}{x_1} + {h_1}{y_1}) - ({k_2}{x_2} + {h_2}{y_2})- ({q_1} + {q_2})z, & \text{for abiotic resources.}
\end{cases}
\end{array}
\label{eq:18}
\end{equation}
\end{widetext}
In both Model B and Model C, two consumer species can coexist either steadily (Fig.S5a, c) or with sustained oscillations (Fig.S5b, d) when there is only one type of resource species. 

\begin{figure*}[ht!]
\includegraphics[width=0.7\linewidth]{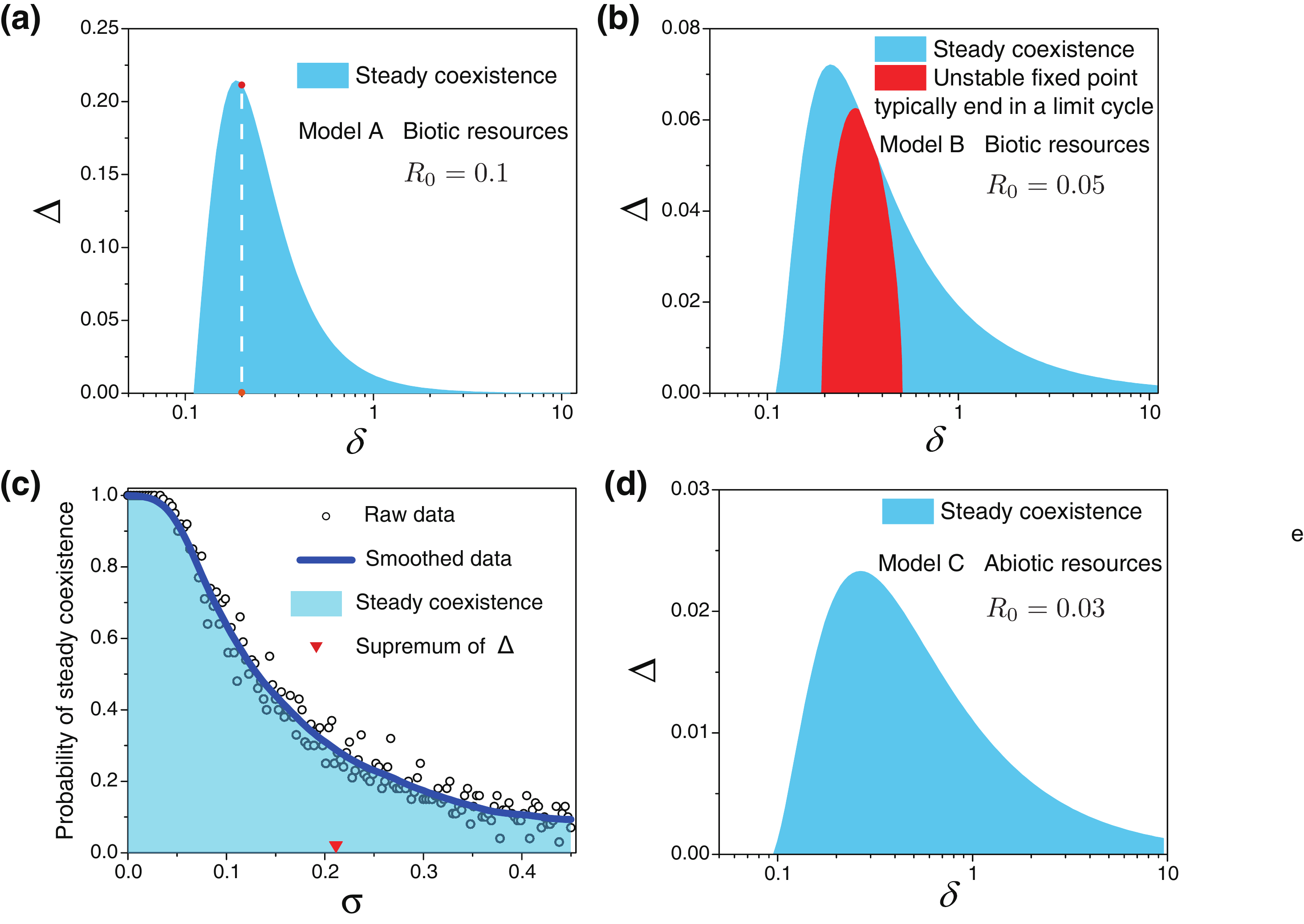}%
\caption{Stable coexistence region of two consumer species competing for one type of resources.
(a), (b), (d) The region of stable coexistence (shown in blue, for globally attracting fixed point) for parameter set. Here $D_i$ ($i$=1, 2) is the only different parameter between consumer species $C_1$ and $C_2$, and $ \Delta \equiv ({D_1} - {D_2})/{D_2}$, the relative difference in mortality rate, measures the competitive differences between the two consumer species. $\delta$ is a dimensionless multiplier that to tune the capture rate and escape rate parameters for the two-consumer species in each scenario. 
(a) Model A, biotic resource. Parameter values at the orange dot is used in calculating all results shown in (c), red dot marks the upper bound of $ \Delta $ that permits species coexistence.
(b) Model B, biotic resource. Red region corresponds to unstable fixed point, which typically ends in a limit cycle (oscillating time series).
(c) Probability of steady coexistence for random parameters. 
First, we chose all parameter exactly the same for two consumer species: $K_0=10$, $D_i=0.005$, $a_i=b_i=0.1$, $k_i=h_i=0.1$, $d_i=e_i= l_i= 0.5$, $w_i=0.1$, $R_0=0.1$ ($i=1,2$). Then, each parameter except $K_0$, $D_2$ (two reduceable parameters, see SI Sec VII for details) was multiplied by a random number following normal distribution $\mathcal{N}(1,\sigma^2)$. All the dots are the raw simulation data (from a sample size of 100), while the line are smoothed data over 25 dots. The blue region corresponds to steady coexistence in the samplings. The inverted red triangle marks the supremum of $\Delta$ 
in (a). 
(d) Model C, abiotic resource.
(a) and (c) were calculated from Eqs.\ref{eq:5}-\ref{eq:6}; (b) was calculated from Eqs.\ref{eq:14}-\ref{eq:15}; (d) was calculated from Eqs.\ref{eq:17}-\ref{eq:18}.
In (a), (b) and (d), we choose an initial set of parameter values for capture rates and escape rates wherever applicable: $k_i^{(0)}=h_i^{(0)}=q_i^{(0)}=0.1$, $d_i^{(0)}=e_i^{(0)}= l_i^{(0)}=s_i^{(0)}= t^{(0)}=0.5$ ($i=1,2$), and then tune those parameters with the multiplier $\delta$ as follows:
$k_i = \delta k_i^{(0)}$,
$h_i = \delta h_i^{(0)}$,
$q_i = \delta h_i^{(0)}$,
$d_i = \delta d_i^{(0)}$,
$e_i = \delta e_i^{(0)}$,
$l_i = \delta l_i^{(0)}$, 
$s_i = \delta s_i^{(0)}$, 
$t = \delta t^{(0)}$ ($i=1,2$).
For other parameters: in (a), (b), (d), $D_2=0.005$, $K_0=10$, $a_i=0.1$, $w_i=0.1$ ($i$=1, 2); in (a), $b_i=0.1$, $R_0=0.1$; in (b), $p_i=0.1$, $R_0=0.05$; in (d), $b_i=0.1$, $p_i=0.1$, $R_0=0.03$. 
}
\label{fig:9}
\end{figure*}
\subsection{\label{sec:level2} Intuitive explanation of why forming chasing triplet can break the CEP}

Intuitively, the reason that forming chasing triplet can break the CEP can be understood from the functional forms of population dynamics at steady state. 
In the classical proof of CEP, in the case of $M=2$ and $N=1$ (Fig.\ref{fig:1}a-c), if both consumers species can coexist at steady state, the abundance of the resource species $R$ needs to satisfy two equations (${f_i}(R)/{D_i} = 1$ ($i$=1, 2)) simultaneously. 
This is equivalent to requiring that two parallel planes share a common point, which is typically impossible (Fig.\ref{fig:7}a). 
In the presence of chasing pairs, as shown in Fig.\ref{fig:7}b, the requirement for steady coexistence corresponds to parallel surfaces (${f_i}({R^{{\rm{(F)}}}})/{D_i} = 1$ ($i$=1, 2), see SI Sec.IV-V for details). 
In the presence of both chasing pairs and chasing triplets, as shown in Fig.\ref{fig:7}c, the requirement for steady coexistence corresponds to three non-parallel surfaces (${\Omega _i}\left( {R,{C_1},{C_2}} \right) = {D_i}{C_i} $($i$=1, 2), $g\left( {R,{C_1},{C_2}} \right)$, see SI Sec.V for details) to cross at one point, which can in principle happen and hence break the CEP. 

\begin{figure*}[ht!]
\includegraphics[width=0.99\linewidth]{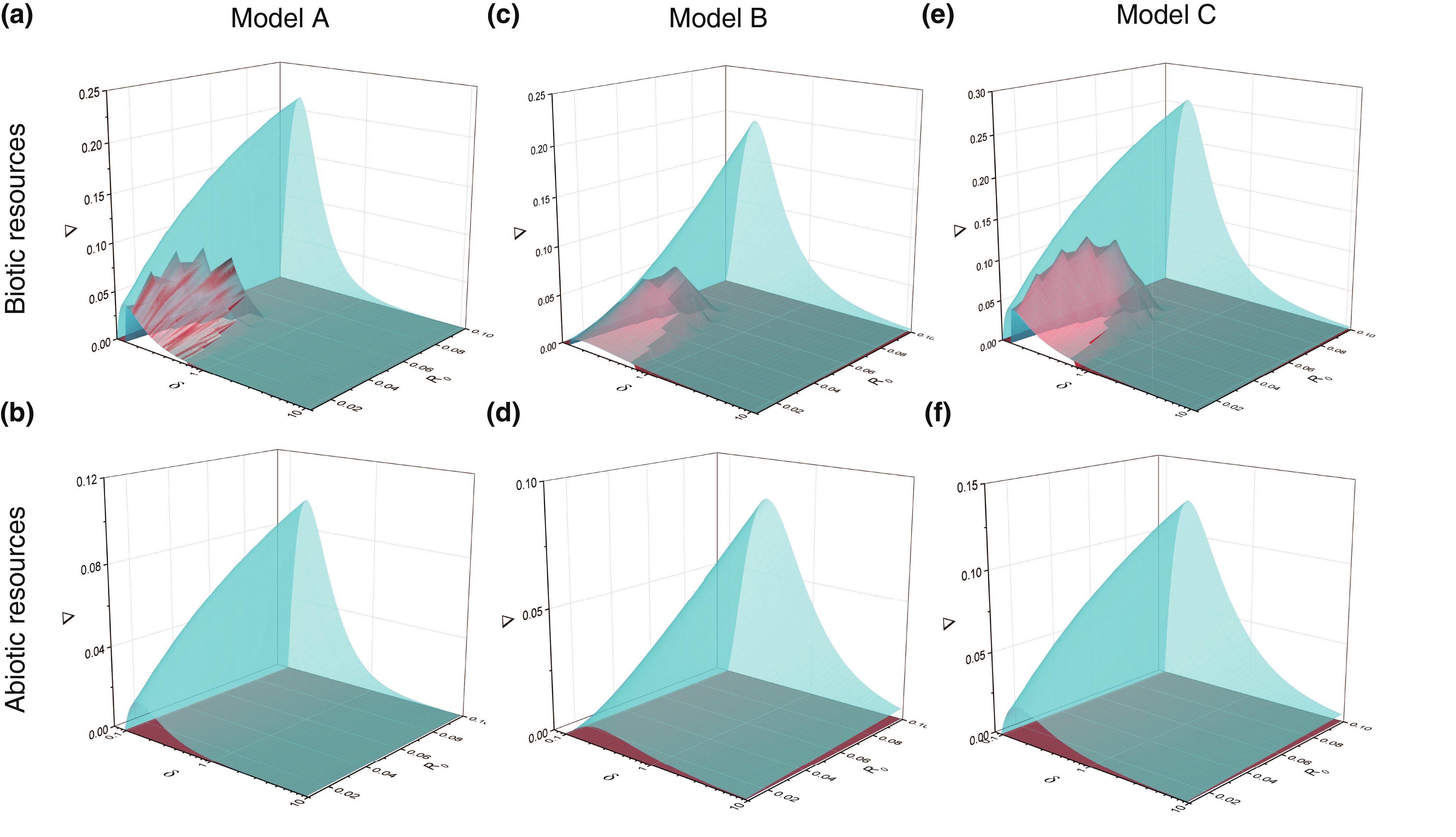}%
\caption{Parameter space of stable coexistence region of two consumer species competing for one type of resources.
Parameters space below the blue surface and above the red surface (in z-axis, for values of $\Delta$) is the stable coexistence region of each case.
Parameters space below the red surface is the unstable fixed-point region, which typically ends in a limit cycle.
(a), (c), (e) Cases of Biotic resources.
(b), (d), (f) Cases of abiotic resources.
(a), (b) Results of Model A.
(c), (d) Results of Model B.
(e), (f) Results of Model C.
(a)-(b) were calculated from Eqs.\ref{eq:5}-\ref{eq:6}; 
(c)-(d) were calculated from Eqs.\ref{eq:14}-\ref{eq:15}; 
(e)-(f) were calculated from Eqs.\ref{eq:17}-\ref{eq:18}.
In (a)-(f), we choose an initial set of parameter values for capture rates and escape rates wherever applicable: $k_i^{(0)}=h_i^{(0)}=q_i^{(0)}=0.1$, $d_i^{(0)}=e_i^{(0)}= l_i^{(0)}=s_i^{(0)}= t^{(0)}=0.5$ ($i=1,2$), and then tune those parameters with the multiplier $\delta$ as follows:
$k_i = \delta k_i^{(0)}$,
$h_i = \delta h_i^{(0)}$,
$q_i = \delta h_i^{(0)}$,
$d_i = \delta d_i^{(0)}$,
$e_i = \delta e_i^{(0)}$,
$l_i = \delta l_i^{(0)}$, 
$s_i = \delta s_i^{(0)}$, 
$t = \delta t^{(0)}$ ($i=1,2$).
For other parameters, in (a), (c) and (e), we choose the following parameters wherever applicable: $D_2=0.005$, $K_0=10$, $a_i=b_i=p_i=0.1$, $w_i=0.1$ ($i$=1, 2); in (b), (d) and (f), we choose the following parameters wherever applicable: $D_2=0.004$, $K_0=5$, $a_i=b_i=p_i=0.1$, $w_i=0.1$ ($i$=1, 2);
}
\label{fig:10}
\end{figure*}
To verify the intuitive explanation, we resort to numerical solutions. Fig.\ref{fig:8} (for biotic resource case) and Fig.S7 (for abiotic resource case) show the numerical results, where the yellow, green and blue surfaces are the exact solutions. 
The parallel green and blue surfaces in the cases of only chasing pair are verified with Fig.\ref{fig:8}a, while the three non-parallel surfaces in scenarios involving chasing triplet are verified with Fig.\ref{fig:8}b-f and Fig.S7. Among the scenarios involving chasing triplet, the fixed points are stable and globally attracting for cases shown in Fig.\ref{fig:8}b (Model A with biotic resources), Fig.\ref{fig:8}c (Model B with biotic resources) and Fig.S7 (Model A with abiotic resources). On the other hand, the presence of chasing triplet does not guarantee species steady coexistence. The non-parallel surfaces may not own a common point in the feasible region (Fig.\ref{fig:8}d) and the fixed point might be unstable (Fig.\ref{fig:8}e-f), which can end in an oscillating coexistence (see Fig.\ref{fig:8}e and the time series in Fig.S4b) or one consumer species dies out (see Fig.\ref{fig:8}f and the time series in Fig.S8b).

Actually, competitive exclusion (i.e., $M \le N$ at steady state) in the classical proof of CEP or the scenario involving only chasing pairs stems from the symmetry constraint of the equation form. In those scenarios, for $M=2$ and $N=1$, there is existence of variable $U \equiv U\left( {R,{C_1},{C_2}} \right)$ satisfying that $\Theta_i \left( {U\left( {R,{C_1},{C_2}} \right)} \right)=D_i$ ($i=1,2$, where $\Theta_i$ is a unspecified function) for the steady-state population dynamics. For the classical proof of CEP, $U=R$ (see Eq.\ref{eq:1}); for the scenario involving only chasing pairs, $U=R^{(\rm{F})}$ (see Eqs.\ref{eq:4} and S13). The existence of $U$ directly leads to parallel planes/surfaces (see Fig.\ref{fig:7}a-b, Fig.8a and Fig. S6b) and thus precludes consumer species coexistence. However, scenario involving chasing triplet or even higher order terms (e.g., quadruplets, quintuplets) breaks the symmetry constraint in the equation form so that there are no existence of such variable $U \equiv U\left( {R,{C_1},{C_2}} \right)$ (otherwise, there cannot be any intersection points in Fig.\ref{fig:8}b-f or Fig.S7, see SI Sec V.A.2 for details). This symmetry breaking enables the breaking of the CEP.

\subsection{\label{sec:level2} Non-special parameter space for species coexistence}

To figure out if there is a non-zero measure parameter space to facilitate species coexistence, we set $D_i$ ($i$=1, 2) to be the only different parameter between consumer species $C_1$ and $C_2$, and all capture rates and escape rates are multiplied by $\delta$ (a dimensionless multiplier) (see Fig.\ref{fig:9}a, b, d). In all Models (Models A-C), for a wide range of $\delta$, we find that there is upper bound tolerance for $\Delta$ (Fig.\ref{fig:9}a, b, d), below which there are coexistence solutions for the two consumer species (the colored region). In the case that the resources are abiotic or for some parameters of $R_0$, the colored region all corresponds to stable coexistence (Fig.\ref{fig:9}a, d, blue region), while in the case that the resources are biotic, for some other parameters of $R_0$, there is a region corresponds to unstable fixed point (Fig.\ref{fig:9}b, red region), which typically ends in a limit cycle. To demonstrate that species coexistence under a non-zero competitive difference (i.e., $\Delta>0$ when $D_i$ is the only different parameter between two consumer species) really means a non-zero parameter space and the supremum of $\Delta>0$ actually measures the likelihood for coexistence, we conducted random sampling analysis. Specifically, we first chose all parameter exactly the same for two consumer species (corresponds to the orange dot in Fig.\ref{fig:9}a). Then, all parameters except $K_0$, $D_2$ (two reducible parameters with dimensionless analysis whose values can be set as arbitrary positive values, see SI Sec VII for details) are multiplied by a random number following normal distribution $\mathcal{N}(1,\sigma^2)$. Obviously, $\sigma$ measures the random extent of the parameter and for each value of $\sigma$, we counted the steady coexistence percentage. The probability of steady coexistence for different values of $\sigma$ is shown in Fig.\ref{fig:9}c, the inverted red triangle denotes the supremum of $\Delta$ for species coexistence, which corresponds to the red dot in Fig.\ref{fig:9}a. When $\sigma$ is small ($\sigma \approx 0$), the probability of steady coexistence is 1, while this probability drops with increasing $\sigma$. When $\sigma=0.45$, this probability approaches 0.1, and the supremum coexistence point of $\Delta$, corresponds to a probability about 0.3. Obviously, $\Delta>0$ (when $D_i$ is the only different parameter between two consumer species) demonstrates a non-zero measure parameter space for coexistence and the value of $\Delta>0$ manifests the likelihood for coexistence.

As shown in Fig.\ref{fig:9}, $R_0$, the growth rate for biotic resource or the supply rate for abiotic resource, might play a critical role for the stability of the fixed point. To further demonstrate this point, we systematically studied the parameter space for stable coexistence. The results are shown in Fig.\ref{fig:10}. Basically, scenarios involving different scenarios of chasing triplets (Model A-C) have qualitatively similar behavior. Here, the region below the blue surface and above the red surface are stable coexistence region, while the region below the red surface and above $\Delta=0$ are the region for unstable fixed point. 
For abiotic resource cases, all fixed points are globally attracting and thus stable (Fig.\ref{fig:10}b, d, f). 
For biotic resource cases, when the value of $R_0$ is small, there is a unstable fixed point region, where trajectories typically end in a limit cycle; when the value of $R_0$ is large, all fixed point are stable (Fig.\ref{fig:10}a, c, e). 
Importantly, there is a non-zero parameter region for all models (Model A-C, biotic or abiotic resources) where the two consumer species can stably coexist (below the blue surface and above the red surface, Fig.\ref{fig:10}), which clearly demonstrates that the violation of
CEP is not due to a special set of model parameters. Note that the violation of CEP in the case of $N=1$ actually implies that it will be violated for more general cases with $N>1$ (see SI Sec.IV for
details).

\section{\label{sec:level1}Discussion}

The CEP has been proposed for several decades. Various mechanisms have been proposed to overcome the constraint set by this principle. Yet, no prior mechanism can generically break the constraint at steady state. Here, by considering the details of the consumption process, especially the possibility to form chasing triplet, our mechanism breaks the original constraint of the CEP. Furthermore, we identify that triplet (or higher order terms) lead to symmetry breaking in the equation form so that there are non-special parameter sets (of non-zero measure) that break the CEP in all scenarios involving different forms of chasing triplet. Meanwhile, we notice that breaking CEP is parameter dependent: for certain parameters, there is no feasible fixed point for coexistence, or the fixed point can be unstable (for biotic resource cases), which may end in a limit cycle. 

The coexistence predicted in our model is testable in experiments, as long as it lies in the stable coexistence region. For biotic resources, or predator-prey system, bacterial predators and their target microbes are potential candidates of consumer and resource species. Yet, $R_0$ is the intrinsic growth rate of resource species in this case, which is hard to be tuned. For abiotic resources, $R_0$ represents the supply rate, which is tunable, then it is possible to see the coexistence or non-coexistence phenomenon by changing $R_0$ (for consumer species of certain range of competitive differences). Actually, all microbial system are potential candidates in this case, yet clear demonstration on the experimental side will be challenging, since it involves disentangling confounding effects such as cross-feeding~\cite{RN515}.

Admittedly, previous mechanism such as temporal or spatial factors, self-organized dynamics, cross-feeding plays significant role in maintaining the biodiversity in nature. Our mechanism based on chasing triplets makes the leap to break the constraint of CEP without making additional assumptions. Our modeling framework is broadly applicable to many consumer-resource ecosystems and hence deepen our understanding of biodiversity in nature.

\section{Acknowledgement}
We thank Babak Momeni, Serguei Saavedra, Chao Tang, Terry Hwa and Nannan Zhao for helpful discussions.
\section{Author Contributions}
X.W and Y.-Y.L conceived and designed the project, developed the model, and wrote
the paper. X.W. carried out all the analytical and numerical calculations.

\begin{widetext}
\newpage
\textbf{\LARGE{Supplementary Information}}

\renewcommand \thesection {\Roman{section}}
\renewcommand \thesubsection {\Alph{subsection}}
\renewcommand \thesubsubsection {\arabic{subsubsection}}

\makeatletter
\newcommand\xleftrightarrow[2][]{%
\ext@arrow 9999{\longleftrightarrowfill@}{#1}{#2}}
\newcommand\longleftrightarrowfill@{%
\arrowfill@\longleftarrow\relbar\longrightarrow}
\renewcommand{\thefigure}{S\@arabic\c@figure}
\renewcommand{\theequation}{S\@arabic\c@equation}
\makeatother

\setcounter{section}{0}    
\setcounter{figure}{0}    
\setcounter{equation}{0}
\renewcommand\theequation{S\arabic{equation}}

\title{\ Overcome Competitive Exclusion in Ecosystems\\
\textit{\textmd{Supplementary Information}}\\}

\author{Xin Wang$^{1}$ and Yang-Yu Liu$^{1,2}$}
\affiliation{$^1$ Channing Division of Network Medicine, Brigham and Women's Hospital and Harvard Medical School, Boston, Massachusetts 02115, USA \\ $^2$ Center for Cancer Systems Biology, Dana-Farber Cancer Institute, Boston, Massachusetts 02115, USA}

\date{\today}

\maketitle
\section{Different forms of the Competitive Exclusion Principle (CEP).}
\subsection{The earliest form.}
The earliest form of the CEP~\cite{RN440,RN459,RN441}, or Gause's law, states that complete competitors cannot coexist, meaning that a more advantageous species can dominate a niche over other species. This was explained in Garret Hardin's classical paper~\cite{RN441} and manifested in Darwin's fitness survival~\cite{RN439}: supposing that one species owns a doubling rate of 1.01 while another species owns a doubling rate of 1, mathematically, it is easy to find that $\mathop {\lim }\limits_{t \to \infty } \frac{{{{\rm{2}}^t}}}{{{2^{1.01t}}}} = 0$, which was interpreted as that a small advantage of one species would ultimately result in extinction of all other competing species~\cite{RN441}.

However, the interpretation above heavily depends on the assumption of exponential growth conditions. Without this assumption, a more advantageous species will not dominate, and coexistence is possible. To illustrate this, here we consider two different scenarios, each contains two types of consumers $C_1$ and $C_2$ feeding on two types of resources $R_1$ and $R_2$.

\subsubsection{Microbial ecosystem in a turbidostat.}

In a turbidostat, resources $R_1$ and $R_2$ (which can be two different types of carbon sources) flow continuously into the system with an adjustable dilution rate $D\left( t \right)$ to keep the turbidity (normally the total amount of $C_1$ and $C_2$, i.e., ${C_{tot}} \equiv {C_1} + {C_2}$) constant. Here we assume that the growth rate terms of $C_i$ ($i$=1, 2) follows Holling's type-II functional response~\cite{RN483}. The population dynamics can be written as follows:
\begin{equation}
\label{Eq_S1}
\left\{ \begin{array}{l}
\dot{R_1}=D\left( t \right){r_1} - D\left( t \right){R_1} - \beta _1^{\left( 1 \right)}\alpha _1^{\left( 1 \right)}{C_1}\frac{{{R_1}}}{{{R_1} + K_1^{\left( 1 \right)}}} - \beta _2^{\left( 1 \right)}\alpha _2^{\left( 1 \right)}{C_2}\frac{{{R_1}}}{{{R_1} + K_2^{\left( 1 \right)}}}\\
\dot{R_2}=D\left( t \right){r_2} - D\left( t \right){R_2} - \beta _1^{\left( 2 \right)}\alpha _1^{\left( 2 \right)}{C_1}\frac{{{R_2}}}{{{R_2} + K_1^{\left( 2 \right)}}} - \beta _2^{\left( 2 \right)}\alpha _2^{\left( 2 \right)}{C_2}\frac{{{R_2}}}{{{R_2} + K_2^{\left( 2 \right)}}}\\
\dot{C_1}=\alpha _1^{\left( 1 \right)}{C_1}\frac{{{R_1}}}{{{R_1} + K_1^{\left( 1 \right)}}} + \alpha _1^{\left( 2 \right)}{C_1}\frac{{{R_2}}}{{{R_2} + K_1^{\left( 2 \right)}}} - {D_1}{C_1} - D\left( t \right){C_1}\\
\dot{C_2}=\alpha _2^{\left( 1 \right)}{C_2}\frac{{{R_1}}}{{{R_1} + K_2^{\left( 1 \right)}}} + \alpha _2^{\left( 2 \right)}{C_2}\frac{{{R_2}}}{{{R_2} + K_2^{\left( 2 \right)}}} - {D_2}{C_2} - D\left( t \right){C_2}
\end{array} \right.,
\end{equation}
where $r_i$ ($i$=1, 2) is the quantity of $R_i$ per unit of the flux into the system; $D_j$ ($j$=1, 2) is the death rate of species $C_j$, while $\alpha _j^{\left( i \right)}$, $\beta _j^{\left( i \right)}$ and $K_j^{\left( i \right)}$ ($i$, $j$=1, 2) are other relevant parameters. Since the turbidity (${C_{tot}} \equiv {C_1} + {C_2}$) is a constant, consider the case that $C_{tot}$ is very small so that $\beta _j^{\left( i \right)}\alpha _j^{\left( i \right)}{C_i}\frac{{{R_i}}}{{{R_i} + K_j^{\left( i \right)}}} \ll D\left( t \right){R_i}$, then $D\left( t \right){r_i} \approx D\left( t \right){R_i}$ (and thus ${r_i} \approx {R_i}$), and the population dynamics of $C_j$ follows:

\begin{equation}
\label{Eq_S2}
\left\{ \begin{array}{l}
\dot{C_1}={C_1}\left( {\frac{{\alpha _1^{\left( 1 \right)}{r_1}}}{{{r_1} + K_1^{\left( 1 \right)}}} + \frac{{\alpha _1^{\left( 2 \right)}{r_2}}}{{{r_2} + K_1^{\left( 2 \right)}}} - {D_1}} \right) - D\left( t \right){C_1}\\
\dot{C_2}={C_2}\left( {\frac{{\alpha _2^{\left( 1 \right)}{r_1}}}{{{r_1} + K_2^{\left( 1 \right)}}} + \frac{{\alpha _2^{\left( 2 \right)}{r_2}}}{{{r_2} + K_2^{\left( 2 \right)}}} - {D_2}} \right) - D\left( t \right){C_2}
\end{array} \right..
\end{equation}

Note that ${\gamma _j} \equiv \sum\limits_{i = 1,2} {\frac{{\alpha _j^{\left( i \right)}{r_i}}}{{{r_i} + K_j^{\left( i \right)}}}} - {D_j}$, the effective growth rate of $C_j$ is fixed once parameters $\alpha _j^{\left( i \right)}$, $K_j^{\left( i \right)}$, $r_i$ and $D_j$ are chosen. Actually, $C_j$ is of exponential growth with effective growth rate $\gamma _j$ and dilution rate $D\left( t \right)$. If ${\gamma _1} > {\gamma _2}$, $\mathop {\lim }\limits_{t \to \infty } D\left( t \right) = {\gamma _1}$ and $\mathop {\lim }\limits_{t \to \infty } \frac{{{C_2}\left( t \right)}}{{{C_1}\left( t \right)}} = 0$, the less competitive species $C_2$ would ultimately become extinct. Overall, in an idealized turbidostat, an advantageous species outcompetes all other species and dominate the system.

\subsubsection{Ecosystem in a natural habitat.}

We consider a natural system with the population dynamics following the classical consumer-resource model~\cite{RN467,RN468}:
\begin{equation}
\label{Eq_S3}
\left\{ \begin{array}{l}
\dot{C_1} = {C_1}\left( {\alpha _1^{\left( 1 \right)}{R_1} + \alpha _1^{\left( 2 \right)}{R_2} - {D_1}} \right)\\
\dot{C_2} = {C_2}\left( {\alpha _2^{\left( 1 \right)}{R_1} + \alpha _2^{\left( 2 \right)}{R_2} - {D_2}} \right)\\
\dot{R_1} = {g_1}\left( {{R_1}} \right) - \beta _1^{\left( 1 \right)}\alpha _1^{\left( 1 \right)}{C_1}{R_1} - \beta _2^{\left( 1 \right)}\alpha _2^{\left( 1 \right)}{C_2}{R_1}\\
\dot{R_2} = {g_2}\left( {{R_1}} \right) - \beta _1^{\left( 2 \right)}\alpha _1^{\left( 2 \right)}{C_1}{R_2} - \beta _2^{\left( 2 \right)}\alpha _2^{\left( 2 \right)}{C_2}{R_2}
\end{array} \right.,
\end{equation}
where $D_j$ ($j$=1, 2) is the death rate of species $C_j$; $\alpha _j^{\left( i \right)}$ and $\beta _j^{\left( i \right)}$ ($i$, $j$=1,2) are other relevant parameters. ${g_i}\left( {{R_i}} \right)$ ($i$=1, 2) is the influx of $R_i$ into the system, where we consider ${g_i}\left( {{R_i}} \right) = {c_i}$ when $R_i$ is abiotic~\cite{RN425} (denote as case A) and ${g_i}\left( {{R_i}} \right) = {R_i}\left[ {{r_i} - {R_i}/K} \right]$ ($r_i$ and $K_i$ are parameters) when $R_i$ is biotic~\cite{RN467} (denote as case B). In both cases, rather than the advantageous species excludes the other, $C_1$ and $C_2$ actually can coexist. To illustrate this, here we consider a simple scheme that $C_1$ and $C_2$ only feed on $R_1$ and $R_2$, respectively, i.e., $\alpha _1^{\left( 2 \right)} = \alpha _2^{\left( 1 \right)} = 0$. Then
\begin{equation}
\label{Eq_S4}
\left\{ \begin{array}{l}
\dot{C_1} = {C_1}\left( {\alpha _1^{\left( 1 \right)}{R_1} - {D_1}} \right)\\
\dot{C_2} = {C_2}\left( {\alpha _2^{\left( 2 \right)}{R_2} - {D_2}} \right)\\
\dot{R_1} = {g_1}\left( {{R_1}} \right) - \beta _1^{\left( 1 \right)}\alpha _1^{\left( 1 \right)}{C_1}{R_1}\\
\dot{R_2} = {g_2}\left( {{R_1}} \right) - \beta _2^{\left( 2 \right)}\alpha _2^{\left( 2 \right)}{C_2}{R_2}
\end{array} \right..
\end{equation}
In case A, at steady state, ${R_i} = {D_i}/\alpha _i^{\left( i \right)}$ ($i$=1, 2) and ${C_i} = {c_i}/\left( {\beta _i^{\left( i \right)}{D_i}} \right)$. In case B, ${R_i} = {D_i}/\alpha _i^{\left( i \right)}$ and ${C_i} = \frac{{{r_i} - {D_i}/\left( {\alpha _i^{\left( i \right)}{K_i}} \right)}}{{\beta _i^{\left( i \right)}\alpha _i^{\left( i \right)}}}$. Since $\alpha _i^{\left( i \right)},\beta _i^{\left( i \right)} > 0$, when ${r_i} > {D_i}/\left( {\alpha _i^{\left( i \right)}{K_i}} \right)$, even if $\alpha _1^{\left( 1 \right)}/{D_1} \gg \alpha _2^{\left( 2 \right)}/{D_2}$ or vice versa, $C_1$ and $C_2$ can coexist in both cases A and B.

Generally, in cases described in Eq.\ref{Eq_S3}, if $\alpha _1^{\left( 1 \right)}/{D_1} > \alpha _2^{\left( 1 \right)}/{D_2}$, as long as $\alpha _1^{\left( 2 \right)}/{D_1} < \alpha _2^{\left( 2 \right)}/{D_2}$, $C_1$ and $C_2$ may coexist. The phenomenon of coexistence can be interpreted as follows: $\alpha _1^{\left( 1 \right)}/{D_1} > \alpha _2^{\left( 1 \right)}/{D_2}$ means that $C_1$ is more advantageous in the competition for $R_1$, while $\alpha _1^{\left( 2 \right)}/{D_1} < \alpha _2^{\left( 2 \right)}/{D_2}$ means that $C_2$ is more advantageous in the competition for $R_2$. Yet, why $C_1$ and $C_2$ can coexist rather the overall advantageous species (the one with larger growth rate) excludes the other?

The underlying reason is that in a long term the growth of $C_1$ and $C_2$ in the competition deviate severely from exponential growth. To illustrate this point, we consider the scenario described in Eq.\ref{Eq_S4} ($\alpha _1^{\left( 2 \right)} = \alpha _2^{\left( 1 \right)} = 0$) with a special case that ignoring the death rate (${D_1} = {D_2} = 0$). In a long term, ${g_i}\left( {{R_i}} \right) = \beta _i^{\left( i \right)}\alpha _i^{\left( i \right)}{C_i}{R_i}$. In case A, $\dot{C_i} = \frac{{{g_i}\left( {{R_i}} \right)}}{{\beta _i^{\left( i \right)}}} = \frac{{{c_i}}}{{\beta _i^{\left( i \right)}}}$, then ${C_i}\left( t \right) = {C_i}\left( {t = 0} \right) + \frac{{{c_i}}}{{\beta _i^{\left( i \right)}}}t$, and $\mathop {\lim }\limits_{t \to \infty } \frac{{{C_2}\left( t \right)}}{{{C_1}\left( t \right)}} = \frac{{{c_2}/\beta _2^{\left( 2 \right)}}}{{{c_1}/\beta _1^{\left( 1 \right)}}}$, where the consumer populations increase linearly rather than exponentially with time. In case B, ${R_i} = {K_i}\left( {{r_i} - \beta _i^{\left( i \right)}\alpha _i^{\left( i \right)}{C_i}} \right)$ and $\dot{C_i} = \alpha _i^{\left( i \right)}{K_i}\left( {{r_i} - \beta _i^{\left( i \right)}\alpha _i^{\left( i \right)}{C_i}} \right){C_i}$, then $\mathop {\lim }\limits_{t \to \infty } {C_i}\left( t \right) = {r_i}/\left( {\beta _i^{\left( i \right)}\alpha _i^{\left( i \right)}} \right)$, where the growth of consumer population is limited by the availability of resources. In both cases, both consumer species are not of exponential growth, and clearly, they do coexist.
\subsection{The classical form since the 1960s.}
Since the 1960s, Robert H. MacArthur and his colleagues~\cite{RN445} formulate the classical form of CEP~\cite{RN445,RN446,RN444}: Consider $M$ types of consumer species competing for $N$ types of resources. At steady state the number of coexisting species of consumers cannot exceed that of resources, i.e., $M \ge N$. This classical CEP form stimulates myriads of studies and is the focus of this paper.
\section{Existing studies.}
\subsection{Existing mechanisms overcome the limit set by the CEP.}
Identifying mechanisms that maintain biodiversity is a central aim in ecology. Various mechanisms~\cite{RN449} have been proposed to overcome the limit set by the CEP and hence explain biodiversity in ecosystems. Those mechanisms can be classified as follows:

a) The ecosystems never reach steady state due to temporal effects of the environment~\cite{RN442,RN461,RN462}: The relaxation time for the system to reach equilibrium is not short enough compared to the frequency of changes in the environment, such as weather, temperature or seasonal cycle.

b) Spatial heterogeneity or patchiness~\cite{RN463,RN460}: Each local patch obeys CEP, while globally support more species of consumers than resource (because there can be a larger overlap of resource species than that of consumers among different patches).

c) Self-organized dynamics promote biodiversity: when the environment remains constant, biodiversity can naturally emerge when the consumers' densities are intrinsically fluctuating~\cite{RN443,RN451} or in a chaos~\cite{RN451,RN452}.

d) Special sets of model parameters (with Lebesgue zero-measure): the simplest example for coexistence of unlimited number of consumers is that all species of consumers share the same ratio of hunting rate to death rate~\cite{RN440}. A recent study~\cite{RN425} found that metabolic trade-offs promote diversity at steady state, but the model heavily relies on the assumption that all consumer species share the same death rate.

e) The biodiversity is facilitated by additional factors other than resources: such as cross-feeding~\cite{RN485,RN486,RN515}, toxin~\cite{RN398}, rock-paper-scissors relation~\cite{RN454,RN464,RN456}, kill the winner~\cite{RN484}, complex interactions~\cite{RN464,RN465,RN456} or co-evolution~\cite{RN453}.
\subsection{GLV models implicitly imply no less resources species than consumers.}
We notice that the Generalized Lotka-Volterra (GLV) model is a very popular modeling framework in the study of biodiversity~\cite{RN469}. However, we emphasize that the GLV model is within the classical constraint of CEP, because it implicitly assumes more (or at least equal number) species of resources than that of the consumers.

Consider the simplest case of two competing species:
\begin{equation}
\label{Eq_S5}
\left\{ \begin{array}{l}
\dot{C_1} = {C_1}\left( {{\alpha _1} - {\beta _{11}}{C_1} - {\beta _{12}}{C_2}} \right)\\
\dot{C_2} = {C_2}\left( {{\alpha _2} - {\beta _{21}}{C_1} - {\beta _{22}}{C_2}} \right)
\end{array} \right..
\end{equation}
Here $C_i$ ($i$=1, 2) stands for the population of consumer species $i$; $\alpha_i$ denotes the growth rate; $\beta_{ij}$ ($i$, $j$=1, 2) denotes the interaction terms. Generally, in GLV models, there is no specific constraint on coefficients $\alpha_i$, $\beta_{ij}$. To clarify the implicit assumption, we consider a consumer-resource model that is comparable to this case with $M=2$ and $N=1$:
\begin{equation}
\label{Eq_S6}
\left\{ \begin{array}{l}
\dot{C_1} = {C_1}\left( {\alpha _1^\prime R - {D_1}} \right)\\
\dot{C_2} = {C_2}\left( {\alpha _2^\prime R - {D_2}} \right)\\
\dot{R} = g\left( {R,{C_1},{C_2}} \right) \equiv {r_R}R\left[ {{r_0}\left( {1 - R/{r_0}} \right) - \beta _1^\prime {C_1} - \beta _2^\prime {C_2}} \right]
\end{array} \right..
\end{equation}
Here, $R$ stands for the population of resources; $\alpha _i^\prime $ ($i$=1, 2) is the growth rate of consumer species $i$; $D_i$ denotes the mortality rate; $g$ follows the classical form of MacArthur's consumer-resource model~\cite{RN467,RN468}. By assuming fast equilibrium for the resource species ($\dot{R}=0$), Eq.\ref{Eq_S6} can be reduced to Eq.\ref{Eq_S5}, with ${\alpha _i} = \alpha _i^\prime {r_0} - {D_i}$, ${\beta _{ij}} = \alpha _i^\prime \beta _j^\prime$ ($i,j$=1, 2). Note that there is a strict constraint on coefficients $\beta_{ij}$: $\frac{{{\beta _{11}}}}{{{\beta _{12}}}} = \frac{{{\beta _{21}}}}{{{\beta _{22}}}}$. With the knowledge of linear algebra~\cite{RN487}, it is easy to prove that only when $M \le N$ can the coefficients in the GLV models be freely chosen.
\subsection{Resources involving chemical compounds.}
Chemical compounds are generally treated as external factors in CEP studies~\cite{RN449}. As shown in Fig.\ref{FigS1}b, there are $N-N'$ ($N>N'$) types of chemical compounds and $N'$ types of normal resources in the ecosystem, while there are $M$ species of consumers. Essentially, within the classical CEP framework, it is permitted that the coexisting $M>N'$ at steady state as long as $M \le N$ (except for special cases corresponding to that shown in Fig.\ref{FigS3}). The proof is same as the schemes shown in Fig.1 of the main text.
\section{Consumption kinetics.}
Consider the simplest scenario of the consumption process, with one type of consumers and one type of resources, i.e., $M=1$ and $N=1$ (Fig.2), and we assume both are biotic. This resembles the simple form of enzymatic reactions,
$${R^{\left( {\rm{F}} \right)}} + {C^{\left( {\rm{F}} \right)}}\xymatrix{
\ar@<0.5ex>[r]^a
& \ar@<0.5ex>[l]^d }
{R^{\left( {\rm{P}} \right)}} \vee {C^{\left( {\rm{P}} \right)}}
\xymatrix{
\ar[r]^k
& }
{C^{\left( {\rm{F}} \right)}}( + ).$$
Here ${C^{\left( {\rm{F}} \right)}}$ and ${R^{\left( {\rm{F}} \right)}}$ stand for the populations of consumers and resources that are freely wandering around, respectively. When a consumer meets a resource with encounter rate $a$, they form a chasing pair ${R^{\left( {\rm{P}} \right)}} \vee {C^{\left( {\rm{P}} \right)}}$ (for simplicity we denote it as $x$). The resource can escape with rate $d$, or be caught and consumed with rate $k$ by the consumer, denoted by ${C^{\left( {\rm{F}} \right)}}( + )$, where '$( + )$' means gaining biomass. By assuming that the transformation process from ${C^{\left( {\rm{F}} \right)}}( + )$ to ${C^{\left( {\rm{F}} \right)}}$ is very fast or consumers can still chase resources when gaining biomass, we count ${C^{\left( {\rm{F}} \right)}}( + )$ as ${C^{\left( {\rm{F}} \right)}}$. By defining the total number of consumers and resources as $C \equiv {C^{\left( {\rm{F}} \right)}} + x$ and $R \equiv {R^{\left( {\rm{F}} \right)}} + x$. The population dynamics of the consumers follows:
\begin{equation}
\label{Eq_S7}
\left\{ \begin{array}{l}
\dot{C}=wkx - {D}C\\
\dot{x} = a{R^{\left( {\rm{F}} \right)}}{C^{\left( {\rm{F}} \right)}} - \left( {d + k} \right)x
\end{array} \right.,
\end{equation}
where $D$ is the mortality rate of consumers (generally ${D} \ll a,k,d$), while the consumption kinetics is given by $kx$. $w$ is a biomass conversion ratio (see maintext Sec II.B).

At steady state $\dot{x} = 0$, rendering a quadratic equation about $x$: ${R^{\left( {\rm{F}} \right)}}{C^{\left( {\rm{F}} \right)}} = \left( {R - x} \right)\left( {C - x} \right) = Kx$, where $K \equiv \frac{{k + d}}{a}$. By considering $0 \le x \le \min \left( {R,C} \right)$, we can easily solve for $x$:
\begin{equation}
\label{Eq_S8}
x = \frac{{\left( {R + C + K} \right)}}{2}\left( {1 - \sqrt {1 - \frac{{4RC}}{{{{\left( {R + C + K} \right)}^2}}}} } \right).
\end{equation}
Since $\frac{{4RC}}{{{{\left( {R + C + K} \right)}^2}}} < 1$, then $\sqrt {1 - \frac{{4RC}}{{{{\left( {R + C + K} \right)}^2}}}} \approx 1 - \frac{{2RC}}{{{{\left( {R + C + K} \right)}^2}}}$, substituting this into Eq.\ref{Eq_S8}, we have $x \approx \frac{{RC}}{{R + C + K}}$ and the consumption kinetics can be approximated as~\cite{RN418}
\begin{equation}
\label{Eq_S9}
wkx \approx \frac{{wkR}}{{R + C + K}}C = f\left( {R,C} \right)C.
\end{equation}
When the consumer population is much smaller than that of resource, i.e., $C \ll R$, the consumption kinetics reduces to the canonical Michaelis-Menten form~\cite{RN43}
\begin{equation}
\label{Eq_S10}
wkx \approx \frac{{wkR}}{{R + K}}C = f\left( R \right)C.
\end{equation}
Note that the $C$-dependency in the growth rate function disappear in the above consumption kinetics. This is also consistent with the growth rate function form $f\left( R \right)$ used in the classical proof of CEP. However, we emphasize that with $C$-dependency in the growth rate functions $f\left( R,C \right)$, the classical proof of CEP does not apply.
\section{Chasing-pair scenarios are under the constraint of CEP}
Although the classical theory does not apply to the $C$-dependent function form $f\left( R,C \right)$, we show below that competitive exclusion principle still holds in the chasing-pair scenarios.

First, we consider the case of $M=2$ and $N=1$ (Fig.3a).
$$
\begin{array}{l}
{R^{\left( {\rm{F}} \right)}} + C_1^{\left( {\rm{F}} \right)}
\xymatrix{
\ar@<0.5ex>[r]^{a_1}
& \ar@<0.5ex>[l]^{d_1} }{R^{\left( {\rm{P}} \right)}} \vee C_1^{\left( {\rm{P}} \right)}
\xymatrix{
\ar[r]^{k_1}
& }
C_1^{\left( {\rm{F}} \right)}( + )\\
{R^{\left( {\rm{F}} \right)}} + C_2^{\left( {\rm{F}} \right)}\xymatrix{
\ar@<0.5ex>[r]^{a_2}
& \ar@<0.5ex>[l]^{d_2}}
{R^{\left( {\rm{P}} \right)}} \vee C_2^{\left( {\rm{P}} \right)}\xymatrix{
\ar[r]^{k_2}
& }
C_2^{\left( {\rm{F}} \right)}( + )
\end{array},
$$
where $C_i^{\left( {\rm{F}} \right)}$ ($i$=1, 2) stands for consumers, $R_j^{\left( {\rm{F}} \right)}$ stands for resources, ${R^{\left( {\rm{P}} \right)}} \vee C_i^{\left( {\rm{P}} \right)}$ (defined as $x_i$) stands for chasing pairs, $C_i^{\left( {\rm{F}} \right)}(+)$ (counted as $C_i^{\left( {\rm{F}} \right)}$) stands for consumers that caught and consumed the resources, $a_i$ stands for encounter rates, $d_i$ stands for escape rates, and $k_i$ stands for capture rates. Denote the total population of consumers and resources at each moment as ${C_i} = C_i^{\left( {\rm{F}} \right)} + {x_i}$ ($i$=1, 2) and $R = {R^{\left( {\rm{F}} \right)}} + {x_1} + {x_2}$. The population dynamics of the consumers and resources can be written as follows:
\begin{equation}
\label{Eq_S11}
\left\{ \begin{array}{l}
\dot{x_1}={a_1}{R^{\left( {\rm{F}} \right)}}C_1^{\left( {\rm{F}} \right)} - \left( {{d_1} + {k_1}} \right){x_1}\\
\dot{x_2}={a_2}{R^{\left( {\rm{F}} \right)}}C_2^{\left( {\rm{F}} \right)} - \left( {{d_2} + {k_2}} \right){x_2}\\
\dot{C_1} ={w_1}{k_1}{x_1} - {D_1}{C_1}\\
\dot{C_2} = {w_2}{k_2}{x_2} - {D_2}{C_2}\\
\dot{R} = g\left( {R,{x_1},{x_2},{C_1},{C_2}} \right)
\end{array} \right.,
\end{equation}
where the functional form of $g\left( {R,{x_1},{x_2},{C_1},{C_2}} \right)$ is unspecific, $D_1$ and $D_2$ denote the death rate of the two consumer species. $w_1$ and $w_2$ are biomass conversion ratios (see maintext Sec II.B).

At steady state, $\dot{x_i}=0$ , we have
\begin{equation}
\label{Eq_S12}
{x_i} = \frac{{{R^{\left( {\rm{F}} \right)}}}}{{{R^{\left( {\rm{F}} \right)}} + {K_i}}}{C_i} = {f_i}\left( {{R^{\left( {\rm{F}} \right)}}} \right){C_i},
\end{equation}
with ${K_i} \equiv \frac{{{d_i} + {k_i}}}{{{a_i}}}$ ($i$=1, 2). Substitute Eq.\ref {Eq_S12} into the third and fourth equations in Eq.\ref{Eq_S11}, with steady-state condition $\dot{C_i}=0$ ($i$=1, 2), we have
\begin{equation}
\label{Eq_S13}
\left\{ \begin{array}{l}
\left( {{f_1}\left( {{R^{\left( {\rm{F}} \right)}}} \right) - {D_1}} \right){C_1} = 0\\
\left( {{f_2}\left( {{R^{\left( {\rm{F}} \right)}}} \right) - {D_2}} \right){C_2} = 0
\end{array} \right..
\end{equation}
If all consumers can coexist, ${f_i}\left( {{R^{\left( {\rm{F}} \right)}}} \right)/{D_i} = 1$ ($i$=1, 2). These relations are depicted in a 2-dimensional graph (Fig.\ref{FigS6}a). Compare Fig.\ref{FigS6}a with Fig.1b, it is evident that the two types of consumers normally cannot coexist at steady state (except for special cases) for similar reason we discussed in the caption of Fig.1.

Now we consider the case of $M=3$ and $N=2$.
$$
\begin{array}{l}
R_1^{\left( {\rm{F}} \right)} + C_1^{\left( {\rm{F}} \right)}\xymatrix{
\ar@<0.5ex>[r]^{a_1^{(1)}}
& \ar@<0.5ex>[l]^{d_1^{(1)}}}
R_1^{\left( {\rm{P}} \right)} \vee C_1^{\left( {\rm{P}} \right)}\xymatrix{
\ar[r]^{k_1^{(1)}}
& }
C_1^{\left( {\rm{F}} \right)}( + )\\
R_1^{\left( {\rm{F}} \right)} + C_2^{\left( {\rm{F}} \right)}\xymatrix{
\ar@<0.5ex>[r]^{a_2^{(1)}}
& \ar@<0.5ex>[l]^{d_2^{(1)}}}
R_1^{\left( {\rm{P}} \right)} \vee C_2^{\left( {\rm{P}} \right)}\xymatrix{
\ar[r]^{k_2^{(1)}}
& }
C_2^{\left( {\rm{F}} \right)}( + )\\
R_1^{\left( {\rm{F}} \right)} + C_3^{\left( {\rm{F}} \right)}\xymatrix{
\ar@<0.5ex>[r]^{a_3^{(1)}}
& \ar@<0.5ex>[l]^{d_3^{(1)}}}
R_1^{\left( {\rm{P}} \right)} \vee C_3^{\left( {\rm{P}} \right)}\xymatrix{
\ar[r]^{k_3^{(1)}}
& }
C_3^{\left( {\rm{F}} \right)}( + )\\
R_2^{\left( {\rm{F}} \right)} + C_1^{\left( {\rm{F}} \right)}\xymatrix{
\ar@<0.5ex>[r]^{a_1^{(2)}}
& \ar@<0.5ex>[l]^{d_1^{(2)}}}
R_2^{\left( {\rm{P}} \right)} \vee C_1^{\left( {\rm{P}} \right)}\xymatrix{
\ar[r]^{k_1^{(2)}}
& }
C_1^{\left( {\rm{F}} \right)}( + )\\
R_2^{\left( {\rm{F}} \right)} + C_2^{\left( {\rm{F}} \right)}\xymatrix{
\ar@<0.5ex>[r]^{a_2^{(2)}}
& \ar@<0.5ex>[l]^{d_2^{(2)}}}
R_2^{\left( {\rm{P}} \right)} \vee C_2^{\left( {\rm{P}} \right)}\xymatrix{
\ar[r]^{k_2^{(2)}}
& }
C_2^{\left( {\rm{F}} \right)}( + )\\
R_2^{\left( {\rm{F}} \right)} + C_3^{\left( {\rm{F}} \right)}\xymatrix{
\ar@<0.5ex>[r]^{a_3^{(2)}}
& \ar@<0.5ex>[l]^{d_3^{(2)}}}
R_2^{\left( {\rm{P}} \right)} \vee C_3^{\left( {\rm{P}} \right)}\xymatrix{
\ar[r]^{k_3^{(2)}}
& }
C_3^{\left( {\rm{F}} \right)}( + )
\end{array},
$$
where $C_i^{\left( {\rm{F}} \right)}$ ($i$=1, 2, 3) stands for consumers; $R_j^{\left( {\rm{F}} \right)}$ ($j$=1,2) stands for resources; $R_j^{\left( {\rm{P}} \right)} \vee C_i^{\left( {\rm{P}} \right)}$ (denoted as $x_i^{\left( j \right)}$; $i$=1-3; $j$=1,2) stands for chasing pairs; $C_i^{\left( {\rm{F}} \right)}(+)$ (counted as $C_i^{\left( {\rm{F}} \right)}$; $i$ =1-3) stands for consumers caught and consumed the resources, $a_i^{\left( j \right)}$ stands for encounter rates, $d_i^{\left( j \right)}$ stands for escape rates, and $k_i^{\left( j \right)}$ stands for capture rates. Denote ${R_j} = R_j^{\left( {\rm{F}} \right)} + \sum\limits_{i = 1}^3 {x_i^{\left( j \right)}}$ ($j$=1, 2) and ${C_i} = C_i^{\left( {\rm{F}} \right)} + \sum\limits_{j = 1}^2 {x_i^{\left( j \right)}}$ ($i$=1-3), the population dynamics can be written as:
\begin{equation}
\label{Eq_S14}
\left\{ \begin{array}{l}
\dot{x_i^{\left( j \right)} }= a_i^{\left( j \right)}R_j^{\left( {\rm{F}} \right)}C_i^{\left( {\rm{F}} \right)} - \left( {d_i^{\left( j \right)} + k_i^{\left( j \right)}} \right)x_i^{\left( j \right)}\\
\dot {C_i} = \sum\limits_{j = 1}^2 {w_i^{\left( j \right)}k_i^{\left( j \right)}x_i^{\left( j \right)}} - {D_i}{C_i}\\
\dot {R_j} = {g_j}\left( {{R_1},{R_2},{C_1},{C_2},{C_3}} \right)
\end{array} \right.,
\end{equation}
with $i$=1-3 and $j$=1, 2. Here the functional form of ${g_j}\left( {{R_1},{R_2},{C_1},{C_2},{C_3}} \right)$ ($j$=1, 2) is unspecific. $D_i$ ($i$=1-3) denotes the death rate of the three consumer species, $w_i^{\left( j \right)}$ are biomass conversion ratios (see maintext Sec II.B).

At steady state, $\dot{x_i^{\left( j \right)}}=0$, we have
\begin{equation}
\label{Eq_S15}
\left\{ \begin{array}{l}
x_i^{\left( 1 \right)} \equiv \frac{{R_1^{\left( {\rm{F}} \right)}}}{{R_2^{\left( {\rm{F}} \right)}K_i^{\left( 1 \right)}/K_i^{\left( 2 \right)} + R_1^{\left( {\rm{F}} \right)} + K_i^{\left( 1 \right)}}}{C_i}\\
x_i^{\left( 2 \right)} \equiv \frac{{R_2^{\left( {\rm{F}} \right)}}}{{R_1^{\left( {\rm{F}} \right)}K_i^{\left( 2 \right)}/K_i^{\left( 1 \right)} + R_2^{\left( {\rm{F}} \right)} + K_i^{\left( 2 \right)}}}{C_i}
\end{array} \right. (i=1\text{-}3),
\end{equation}
where $K_i^{\left( j \right)} \equiv \frac{{d_i^{\left( j \right)} + k_i^{\left( j \right)}}}{{a_i^{\left( j \right)}}}$ ($i$=1-3 ; $j$=1, 2). Hence

$\sum\limits_{j = 1}^2 {w_i^{\left( j \right)}k_i^{\left( j \right)}x_i^{\left( j \right)}} = \left( {\frac{{w_i^{\left( 1 \right)}k_i^{\left( 1 \right)}R_1^{\left( {\rm{F}} \right)}}}{{{{R_2^{\left( {\rm{F}} \right)}K_i^{\left( 1 \right)}} \mathord{\left/
{\vphantom {{R_2^{\left( {\rm{F}} \right)}K_i^{\left( 1 \right)}} {K_i^{\left( 2 \right)}}}} \right.
\kern-\nulldelimiterspace} {K_i^{\left( 2 \right)}}} + R_1^{\left( {\rm{F}} \right)} + K_i^{\left( 1 \right)}}} + \frac{{w_i^{\left( 2 \right)}k_i^{\left( 2 \right)}R_2^{\left( {\rm{F}} \right)}}}{{{{R_1^{\left( {\rm{F}} \right)}K_i^{\left( 2 \right)}} \mathord{\left/
{\vphantom {{R_1^{\left( {\rm{F}} \right)}K_i^{\left( 2 \right)}} {K_i^{\left( 1 \right)}}}} \right.
\kern-\nulldelimiterspace} {K_i^{\left( 1 \right)}}} + R_2^{\left( {\rm{F}} \right)} + K_i^{\left( 2 \right)}}}} \right){C_i} \equiv {f_i}\left( {R_1^{\left( {\rm{F}} \right)},R_2^{\left( {\rm{F}} \right)}} \right){C_i}$.
Substitute the expression of $\sum\limits_{j = 1}^2 {w_i^{\left( j \right)}k_i^{\left( j \right)}x_i^{\left( j \right)}}$ into Eq.\ref{Eq_S14}, with steady-state condition $\dot{C_i}=0$ ($i$=1- 3), we have
\begin{equation}
\label{Eq_S16}
\left\{ \begin{array}{l}
\left( {{f_1}\left( {R_1^{\left( {\rm{F}} \right)},R_2^{\left( {\rm{F}} \right)}} \right) - {D_1}} \right){C_1} = 0\\
\left( {{f_2}\left( {R_1^{\left( {\rm{F}} \right)},R_2^{\left( {\rm{F}} \right)}} \right) - {D_2}} \right){C_2} = 0\\
\left( {{f_3}\left( {R_1^{\left( {\rm{F}} \right)},R_2^{\left( {\rm{F}} \right)}} \right) - {D_3}} \right){C_3} = 0
\end{array} \right..
\end{equation}
If all consumers can coexist, ${f_i}\left( {R_1^{\left( {\rm{F}} \right)},R_2^{\left( {\rm{F}} \right)}} \right) = {D_i}$ ($i$=1-3). These relations are depicted in a plane as shown in Fig.\ref{FigS6}c. Compare Fig.\ref{FigS6}c with Fig.1e, it is evident that the three types of consumers normally cannot all coexist (except for special cases).

This method can be extended to general cases of $M>N$, where we can obtain a general set of equations in the form of Eqs.~\ref{Eq_S13}~and~\ref{Eq_S16}.
\section{Forming chasing triplets can overcome CEP.}
Considering again the consumption process, when a consumer is chasing a resource and forming a chasing pair, other consumers, especially consumers of the same species may join to chase the same resource individual. Consider the case of $M=2$ and $N=1$ (Fig.3a, b, we denote the combination of both scenarios as Model A), but now two consumers of the same species can chase the same resource, forming a chasing triplet (Fig.3b). The consumption process can be described as follows:
$$
\begin{array}{l}
{R^{\left( {\rm{F}} \right)}} + C_1^{\left( {\rm{F}} \right)}\xymatrix{
\ar@<0.5ex>[r]^{a_1}
& \ar@<0.5ex>[l]^{d_1}}
{R^{\left( {\rm{P}} \right)}} \vee C_1^{\left( {\rm{P}} \right)} \xymatrix{
\ar[r]^{k_1}
& }
C_1^{\left( {\rm{F}} \right)}( + )\\
{R^{\left( {\rm{F}} \right)}} + C_2^{\left( {\rm{F}} \right)}\xymatrix{
\ar@<0.5ex>[r]^{a_2}
& \ar@<0.5ex>[l]^{d_2}}
{R^{\left( {\rm{P}} \right)}} \vee C_2^{\left( {\rm{P}} \right)}\xymatrix{
\ar[r]^{k_2}
& }
C_2^{\left( {\rm{F}} \right)}( + )\\
{R^{\left( {\rm{P}} \right)}} \vee C_1^{\left( {\rm{P}} \right)} + C_1^{\left( {\rm{F}} \right)}\xymatrix{
\ar@<0.5ex>[r]^{b_1}
& \ar@<0.5ex>[l]^{e_1}}
C_1^{\left( {\rm{T}} \right)} \vee {R^{\left( {\rm{T}} \right)}} \vee C_1^{\left( {\rm{T}} \right)}\\
{R^{\left( {\rm{P}} \right)}} \vee C_2^{\left( {\rm{P}} \right)} + C_2^{\left( {\rm{F}} \right)}\xymatrix{
\ar@<0.5ex>[r]^{b_2}
& \ar@<0.5ex>[l]^{e_2}}
C_2^{\left( {\rm{T}} \right)} \vee {R^{\left( {\rm{T}} \right)}} \vee C_2^{\left( {\rm{T}} \right)}\\
C_1^{\left( {\rm{T}} \right)} \vee {R^{\left( {\rm{T}} \right)}} \vee C_1^{\left( {\rm{T}} \right)}\xymatrix{
\ar[r]^{h_1}
& }
C_1^{\left( {\rm{F}} \right)}( + ) + C_1^{\left( {\rm{F}} \right)}\\
C_2^{\left( {\rm{T}} \right)} \vee {R^{\left( {\rm{T}} \right)}} \vee C_2^{\left( {\rm{T}} \right)}\xymatrix{
\ar[r]^{h_2}
& }
C_2^{\left( {\rm{F}} \right)}( + ) + C_2^{\left( {\rm{F}} \right)}\\
C_1^{\left( {\rm{T}} \right)} \vee {R^{\left( {\rm{T}} \right)}} \vee C_1^{\left( {\rm{T}} \right)}\xymatrix{
\ar[r]^{l_1}
& }
C_1^{\left( {\rm{F}} \right)} + C_1^{\left( {\rm{F}} \right)} + {R^{\left( {\rm{F}} \right)}}\\
C_2^{\left( {\rm{T}} \right)} \vee {R^{\left( {\rm{T}} \right)}} \vee C_2^{\left( {\rm{T}} \right)}\xymatrix{
\ar[r]^{l_2}
& }
C_2^{\left( {\rm{F}} \right)} + C_2^{\left( {\rm{F}} \right)} + {R^{\left( {\rm{F}} \right)}}
\end{array},
$$
where $C_i^{\left( {\rm{F}} \right)}$ ($i$=1, 2) and $R^{\left( {\rm{F}} \right)}$ stand for freely wandering consumers and resources respectively, ${R^{\left( {\rm{P}} \right)}} \vee C_i^{\left( {\rm{P}} \right)}$ (denoted as $x_i$; $i$=1, 2) stands for chasing pairs, $C_i^{\left( {\rm{T}} \right)} \vee {R^{\left( {\rm{T}} \right)}} \vee C_i^{\left( {\rm{T}} \right)}$ (denoted as $y_i$) stands for chasing triplets, $C_i^{\left( {\rm{F}} \right)}(+)$ (counted as $C_i^{\left( {\rm{F}} \right)}$) stands for consumers caught and consumed the resources, and $a_i$, $b_i$, $d_i$, $e_i$, $h_i$, $k_i$ and $l_i$ stand for relevant parameters specified in Fig.3a-b. Denote $R = {R^{\left( {\rm{F}} \right)}} + \sum\limits_{i = 1}^2 {\left( {{x_i} + {y_i}} \right)}$ and ${C_i} = C_i^{\left( {\rm{F}} \right)} + {x_i} + 2{y_i}$ ($i$=1, 2), the population dynamics can be written as follows:
\begin{equation}
\label{Eq_S17}
\left\{ \begin{array}{l}
\dot{x_1} = {a_1}{R^{\left( {\rm{F}} \right)}}C_1^{\left( {\rm{F}} \right)} - \left( {{d_1} + {k_1}} \right){x_1} - {b_1}{x_1}C_1^{\left( {\rm{F}} \right)} + {e_1}{y_1}\\
\dot {x_2} = {a_2}{R^{\left( {\rm{F}} \right)}}C_2^{\left( {\rm{F}} \right)} - \left( {{d_2} + {k_2}} \right){x_2} - {b_2}{x_2}C_2^{\left( {\rm{F}} \right)} + {e_2}{y_2}\\
\dot {y_1} = {b_1}{x_1}C_1^{\left( {\rm{F}} \right)} - \left( {{h_1} + {e_1} + {l_1}} \right){y_1}\\
\dot {y_2} = {b_2}{x_2}C_2^{\left( {\rm{F}} \right)} - \left( {{h_2} + {e_2} + {l_2}} \right){y_2}\\
\dot {C_1} ={w_1}( {k_1}{x_1} + {h_1}{y_1}) - {D_1}{C_1}\\
\dot {C_2} = {w_2}({k_2}{x_2} + {h_2}{y_2}) - {D_2}{C_2}\\
\dot {R} = g^\prime\left( {R,{x_1},{x_2},{y_1},{y_2},{C_1},{C_2}} \right)
\end{array} \right.,
\end{equation}
where $D_i$ ($i$=1, 2) denotes the death rate of the consumer species. At steady state, $\dot{x_i}=0$, $\dot{y_i}=0$ ($i$=1, 2), we have
\begin{equation}
\label{Eq_S18}
\left\{ \begin{array}{l}
{a_1}{R^{\left( {\rm{F}} \right)}}\left( {{C_1} - {x_1} - 2{y_1}} \right) - \left( {{d_1} + {k_1}} \right){x_1} - {b_1}{x_1}\left( {{C_1} - {x_1} - 2{y_1}} \right) + {e_1}{y_1} = 0\\
{a_2}{R^{\left( {\rm{F}} \right)}}\left( {{C_2} - {x_2} - 2{y_2}} \right) - \left( {{d_2} + {k_2}} \right){x_2} - {b_2}{x_2}\left( {{C_2} - {x_2} - 2{y_2}} \right) + {e_2}{y_2} = 0\\
{b_1}{x_1}\left( {{C_1} - {x_1} - 2{y_1}} \right) - \left( {{h_1} + {e_1} + {l_1}} \right){y_1} = 0\\
{b_2}{x_2}\left( {{C_2} - {x_2} - 2{y_2}} \right) - \left( {{h_2} + {e_2} + {l_2}} \right){y_2} = 0
\end{array} \right..
\end{equation}
Define
\begin{equation}
\label{Eq_S19}
\left\{ \begin{array}{l}
P_1^{\left( i \right)} = \left( {2{d_i} + 2{k_i} - {h_i} - {l_i}} \right){b_i},\\
P_2^{\left( i \right)} = \left( {{h_i} + {e_i} + {l_i}} \right){a_i},\\
P_3^{\left( i \right)} = {{\left( {{d_i} + {k_i}} \right)} \mathord{\left/
{\vphantom {{\left( {{d_i} + {k_i}} \right)} {{a_i}}}} \right.
\kern-\nulldelimiterspace} {{a_i}}},\\
P_4^{\left( i \right)} = \left( {{h_i} + {l_i}} \right){b_i},\\
P_5^{\left( i \right)} = {{\left( {{h_i} + {l_i}} \right)} \mathord{\left/
{\vphantom {{\left( {{h_i} + {l_i}} \right)} {{a_i}}}} \right.
\kern-\nulldelimiterspace} {{a_i}}},
\end{array} \right.\quad i=1, 2.
\end{equation}
From Eq.\ref{Eq_S18}
\begin{equation}
\label{Eq_S20}
P_1^{\left( i \right)}{x_i}^2 + \left[ {P_2^{\left( i \right)}\left( {P_3^{\left( i \right)} + {R^{\left( {\rm{F}} \right)}}} \right) + P_4^{\left( i \right)}{C_1}} \right]{x_i} - P_2^{\left( i \right)}{R^{\left( {\rm{F}} \right)}}{C_1} = 0,
\end{equation}
and 
\begin{equation}
\label{Eq_S21}
{y_i} = \frac{{{R^{\left( {\rm{F}} \right)}}{C_i} - \left( {P_3^{\left( i \right)} + {R^{\left( {\rm{F}} \right)}}} \right){x_i}}}{{2{R^{\left( {\rm{F}} \right)}} + P_5^{\left( i \right)}}},
\end{equation}
with $i$=1, 2. When $P_1^{\left( i \right)}\ne 0$, note that $0 \le {x_i} \le \min \left( {{C_i},R} \right)$, then 
\begin{equation}
\label{Eq_S22}
\left\{ \begin{array}{l}
{x_i} = \frac{\begin{array}{l}
\sqrt {{{\left[ {P_2^{\left( i \right)}\left( {P_3^{\left( i \right)} + {R^{\left( {\rm{F}} \right)}}} \right) + P_4^{\left( i \right)}{C_i}} \right]}^2} + 4P_1^{\left( i \right)}P_2^{\left( i \right)}{R^{\left( {\rm{F}} \right)}}{C_i}} \\
- \left[ {P_2^{\left( i \right)}\left( {P_3^{\left( i \right)} + {R^{\left( {\rm{F}} \right)}}} \right) + P_4^{\left( i \right)}{C_i}} \right]
\end{array}}{{2P_1^{\left( i \right)}}} \equiv u_i^\prime \left( {{R^{\left( {\rm{F}} \right)}},{C_i}} \right)\\
{y_i} = \frac{{{R^{\left( {\rm{F}} \right)}}{C_i} - \left( {P_3^{\left( i \right)} + {R^{\left( {\rm{F}} \right)}}} \right)u_i^\prime \left( {{R^{\left( {\rm{F}} \right)}},{C_i}} \right)}}{{2{R^{\left( {\rm{F}} \right)}} + P_5^{\left( i \right)}}} \equiv v_i^\prime \left( {{R^{\left( {\rm{F}} \right)}},{C_i}} \right)
\end{array} \right. (i=1, 2).
\end{equation}
Note that ${R^{\left( {\rm{F}} \right)}} = R - \sum\limits_{i = 1}^2 {\left( {{x_i} + {y_i}} \right)}$, combined with Eq.\ref{Eq_S18}, we get $x_i$, $y_i$ of the following form:
\begin{equation}
\label{Eq_S23}
\left\{ \begin{array}{l}
{x_i} = {u_i}\left( {R,{C_1},{C_2}} \right)\\
{y_i} = {v_i}\left( {R,{C_1},{C_2}} \right)
\end{array} \right. (i=1,2).
\end{equation}
Consequently,
\begin{equation}
\label{Eq_S24}
{w_i}({k_i}{x_i} + {h_i}{y_i}) \equiv \Omega _i^\prime \left( {{R^{\left( {\rm{F}} \right)}},{C_1},{C_2}} \right) \equiv {\Omega _i}\left( {R,{C_1},{C_2}} \right) (i=1,2).
\end{equation}
Importantly, as long as ${b_i} \ne 0$ ($i$=1, 2), there is no existence of such variable $U \equiv U\left( {R,{C_1},{C_2}} \right)$ that satisfy the equality: $\Theta_i \left( {U\left( {R,{C_1},{C_2}} \right)} \right) = \frac{{{\Omega _i}\left( {R,{C_1},{C_2}} \right)}}{{{C_i}}}$ (where function $\Theta_i$ is unspecific, see SI Sec V.A.2 for details). At steady state, $\dot{C_i}=0$ ($i$=1, 2) and $\dot{R}=0$. Substituting Eqs.~\ref{Eq_S23}-\ref{Eq_S24} into Eq.\ref{Eq_S17}, we get
\begin{equation}
\label{Eq_S25}
\left\{ \begin{array}{l}
{\Omega _1}\left( {R,{C_1},{C_2}} \right) - {D_1}{C_1} = 0\\
{\Omega _2}\left( {R,{C_1},{C_2}} \right) - {D_2}{C_2} = 0\\
g\left( {R,{C_1},{C_2}} \right) = 0
\end{array} \right.,
\end{equation}
where 
$g\left( {R,{C_1},{C_2}} \right) \equiv g'\left( {R,{u_i}\left( {R,{C_1},{C_2}} \right),{v_i}\left( {R,{C_1},{C_2}} \right),{C_1},{C_2}} \right)$. 

\subsection{Intuitive explanation of why forming chasing triplet can break CEP}
\subsubsection{Comparison between the classical case, chasing pair scenario and chasing pair+triplet scenario}
With Eq.\ref{Eq_S25}, we can give an intuitive explanation why forming chasing triplet may break CEP using the functional forms of population dynamics at steady state. To illustrate how the consumers are liberated from the constraint of CEP in the presence of chasing triplets, we compare it with the classical proof scenario described in Eq.1 and the chasing-pair scenario described with Eq.\ref{Eq_S13}, in the case of $M=2$ and $N=1$.

In the classical case (Eq.1), if both consumers can coexist at steady state, ${f_i}\left( R \right)/{D_i} = 1$ ($i$=1, 2). Now we depict these relations in a three-dimensional space as shown in Fig.7a, where $C_1$ is the $x$-axis, $C_2$ the $y$-axis and $R$ the $z$-axis. The green plane corresponds to ${f_1}\left( R \right)/{D_1} = 1$ while the blue plane corresponds to ${f_2}\left( R \right)/{D_2} = 1$. Note that in principle there could be multiple green/blue planes if the equation ${f_i}\left( R \right)/{D_i} = 1$ has multiple solutions. These planes are parallel to the plane $R=0$ and hence do not share a common point (except for special cases).

In the presence of chasing pairs (Eq.\ref{Eq_S13}), if the two consumer species can coexist at steady state, ${f_i}\left( {{R^{\left( {\rm{F}} \right)}}} \right)/{D_i} = 1$ ($i$=1, 2). On one hand, we can depict these relations in Fig.\ref{FigS6}b, where $C_1$ is the $x$-axis, $C_2$ the $y$-axis and ${R^{\left( {\rm{F}} \right)}}$ the $z$-axis. The green plane corresponds to ${f_1}\left( {{R^{\left( {\rm{F}} \right)}}} \right)/{D_1} = 1$ while the blue plane corresponds to ${f_2}\left( {{R^{\left( {\rm{F}} \right)}}} \right)/{D_2} = 1$. Those planes are parallel to the plane ${R^{\left( {\rm{F}} \right)}} = 0$, and thus do not share a common point (except for special cases). On the other hand, we can depict the relations in Eq.\ref{Eq_S13} in a coordinate where the $z$-axis is $R$ rather than ${R^{\left( {\rm{F}} \right)}}$. As shown in Fig.7b, the green surface corresponds to ${f_1}\left( {{R^{\left( {\rm{F}} \right)}}} \right)/{D_1} = 1$ while the blue surface corresponds to ${f_2}\left( {{R^{\left( {\rm{F}} \right)}}} \right)/{D_2} = 1$. Essentially, it is a coordinate transformation from Fig.\ref{FigS6}b. With the knowledge of topology~\cite{RN482}, we know that the green surface is parallel to the blue surface and normally do not share a common point (except for special cases that two surfaces completely overlap).

In the presence of chasing triplets (Eq.\ref{Eq_S25}), we depict the relations in Fig.7c, where $C_1$ is the $x$-axis, $C_2$ the $y$-axis and $R$ the $z$-axis. The green surface corresponds to ${w_1}\left( {R,{C_1},{C_2}} \right) = {D_1}{C_1}$, while the blue surface corresponds to ${w_2}\left( {R,{C_1},{C_2}} \right) = {D_2}{C_2}$, and the yellow surface corresponds to $g\left( {R,{C_1},{C_2}} \right) = 0$. As determined from Eq.\ref{Eq_S17}, the green surface is not parallel to the blue one, and thus they have at least one intersection curve (shown as the dashed purple curve in Fig.7c). Since a curve and a surface can normally have an intersection point, the three surfaces of different colors can normally have at least one intersection point (shown as the red point in Fig.7c). As long as those intersection points locate within the feasible region, i.e., $\min \left( {R,{C_1},{C_2}} \right) > 0$, the two consumer species can coexist at steady state. 
Numerical results (exact solution) shown in Fig.8 (comparable to Fig.7) confirm our intuitive explanation.
\subsubsection{Triplet or higher order terms lead to symmetry breaking in the constraint of the CEP}
The numerical results shown in Fig.8b-c and Fig.\ref{FigS7} clearly demonstrate that in the presence of chasing triplet, the three surfaces that correspond to $\dot{C_1}=0$, $\dot{C_2}=0$ and $\dot{R}=0$ are unparallel to each other and can share an intersect point (red points in Fig.8b-c and Fig.\ref{FigS7}). This means that in Eq.\ref{Eq_S24}, it is impossible for any variable, say $U \equiv U\left( {R,{C_1},{C_2}} \right)$ to satisfy the equality:$\Theta_i \left( {U\left( {R,{C_1},{C_2}} \right)} \right) = \frac{{{\Omega _i}\left( {R,{C_1},{C_2}} \right)}}{{{C_i}}}$ (where function $\Theta_i$ is unspecific). Otherwise, ${\Theta _i}\left( {U\left( {R,{C_1},{C_2}} \right)} \right) = {D_i}$ ($i$=1, 2), the planes that correspond to $\dot{C_1}=0$ and $\dot{C_2}=0$ are parallel to the $C_1$-$O$-$C_2$ plane ($O$ is the origin point) in the $(C_1, C_2, {U\left( {R,{C_1},{C_2}} \right)})$ coordinate and corresponds to parallel surfaces in the $(C_1, C_2, R)$ coordinate. Meanwhile, in the classical case, $R$ corresponds to $U\left( {R,{C_1},{C_2}} \right)$ and in the chasing pair scenario, ${R^{\left( {\rm{F}} \right)}}$ corresponds to $U\left( {R,{C_1},{C_2}} \right)$. 

To investigate why there was no existence of $U( {R,{C_1},{C_2}})$ in scenario involving chasing triplet, we revisit the steady state form of Eq.\ref{Eq_S17}. Combined with ${C_i} = C_i^{\left( {\rm{F}} \right)} + {x_i} + 2{y_i}$ ($i$=1, 2), then 
\begin{equation}
\label{Eq_S26}
\left\{ \begin{array}{l}
C_i^{\left( {\rm{F}} \right)} = \frac{{({C_i} - {x_i})({h_i} + {e_i} + {l_i})}}{{{h_i} + {e_i} + {l_i} + 2{b_i}{x_i}}},\\
{R^{\left( {\rm{F}} \right)}} = \frac{{\left( {{d_i} + {k_i}} \right)}}{{{a_i}}}\frac{{{x_i}\left( {{h_i} + {e_i} + {l_i} + 2{b_i}{x_i}} \right)}}{{({C_i} - {x_i})({h_i} + {e_i} + {l_i})}} + \frac{{{b_i}\left( {{h_i} + {l_i}} \right)}}{{{a_i}\left( {{h_i} + {e_i} + {l_i}} \right)}}{x_i},\\
{y_i} = \frac{{{b_i}}}{{{h_i} + {e_i} + {l_i}}}{x_i}C_i^{\left( {\rm{F}} \right)} = {x_i}\frac{{{b_i}({C_i} - {x_i})}}{{{h_i} + {e_i} + {l_i} + 2{b_i}{x_i}}}.
\end{array} \right. \quad i=1, 2.
\end{equation}
From the last two equations in Eq.\ref{Eq_S26}, we find that
\begin{equation}
\label{Eq_S27}
\left\{ \begin{array}{l}
{x_i} = {x_i}\left( {{R^{\left( {\rm{F}} \right)}}} \right),\\
{y_i}={y_i}\left( {{R^{\left( {\rm{F}} \right)}}}, C_i \right) = {x_i}\left( {{R^{\left( {\rm{F}} \right)}}} \right)\frac{{{b_i}({C_i} - {x_i}\left( {{R^{\left( {\rm{F}} \right)}}} \right))}}{{{h_i} + {e_i} + {l_i} + 2{b_i}{x_i}\left( {{R^{\left( {\rm{F}} \right)}}} \right)}}.
\end{array} \right.
\end{equation}
Then
\begin{equation}
\label{Eq_S28}
\frac{{{\Omega _i}\left( {R,{C_1},{C_2}} \right)}}{{{C_i}}}=\frac{{{w_i}}}{{{C_i}}}\left( {{k_i}{x_i}\left( {{R^{{\rm{(F)}}}}} \right) + {h_i}{y_i}\left( {{R^{{\rm{(F)}}}},{C_i}} \right)} \right) \equiv {\Theta _i^\prime}\left( {{R^{{\rm{(F)}}}},{C_i}} \right), \quad i=1, 2.
\end{equation} 
Note that in Eq.\ref{Eq_S28}, only when $b_i=0$, can ${\Theta _i^\prime}\left( {{R^{{\rm{(F)}}}},{C_i}} \right)$ be reduced to $ {\Theta _i^\prime}\left( {{R^{{\rm{(F)}}}}} \right)$, otherwise there is no existence of $U(R,C_1,C_2)$. Consequently, the triplet term $y_i$ ($b_i \ne 0$) breaks the symmetric constraint in the equations form, i.e., the existence of $U(R,C_1,C_2)$, which overcomes CEP. Similarly, Models B-C or scenarios involving even higher order terms (e.g. quadruplet, quintuplets) are subject to the same analysis above and results in no existence of $U(R,C_1,C_2)$. Actually, chasing pair scenario is a special case of Model A when $b_i=0$, and so does triplet scenario for quadruplet (or quintuplets et.al) scenarios. Thus, the fact that chasing triplet scenario can overcome CEP naturally means that all higher order terms scenarios (triplet or higher) can break CEP. In sum, higher order terms (triplet or higher) lead to symmetry breaking in the constraint of the equation form that overcomes CEP. 
\subsection{Analytical solutions to steady-state species abundances }
Generically, there is no closed form solution to Eq.\ref{Eq_S17}. However, when the abundance of resources are much larger than that of consumers, $R \gg {C_1},{C_2}$, which applies to almost all cases in the wild, then $R \approx {R^{\left( {\rm{F}} \right)}}$. Combining these results with $\dot{C_i}=0$ ($i$=1, 2) and $\dot{g}=0$, 
$$
\left\{ \begin{array}{l}
{x_i} \approx \frac{{\left[ {P_2^{\left( i \right)}\left( {P_3^{\left( i \right)} + R} \right) + P_4^{\left( i \right)}{C_i}} \right]}}{{2P_1^{\left( i \right)}}}\left\{ {\sqrt {1 + \frac{{4P_1^{\left( i \right)}P_2^{\left( i \right)}R{C_i}}}{{{{\left[ {P_2^{\left( i \right)}\left( {P_3^{\left( i \right)} + R} \right) + P_4^{\left( i \right)}{C_i}} \right]}^2}}}} - 1} \right\}\\
{y_i} \approx \frac{{R{C_i} - \left( {P_3^{\left( i \right)} + R} \right){x_i}}}{{2R + P_5^{\left( i \right)}}}
\end{array} \right.,
$$

Note that $\frac{{4P_1^{\left( i \right)}P_2^{\left( i \right)}R{C_i}}}{{{{\left[ {P_2^{\left( i \right)}\left( {P_3^{\left( i \right)} + R} \right) + P_4^{\left( i \right)}{C_i}} \right]}^2}}} \sim \frac{{4P_1^{\left( i \right)}RC}}{{P_2^{\left( i \right)}{{\left( {P_3^{\left( i \right)} + R} \right)}^2}}} \sim \frac{C}{R} \ll 1$, where $\sim$ means the order of magnitude is similar. Then, using the approximation that $\sqrt {1 - x} \approx 1 - x/2$ (when $x \ll 1$), 
$$\left\{ \begin{array}{l}
{x_i} \approx \frac{{P_2^{\left( i \right)}R{C_i}}}{{P_2^{\left( i \right)}\left( {P_3^{\left( i \right)} + R} \right) + P_4^{\left( i \right)}{C_i}}}\\
{y_i} \approx \frac{{R{C_i}}}{{2R + P_5^{\left( i \right)}}}\left[ {\frac{{P_4^{\left( i \right)}{C_i}}}{{P_2^{\left( i \right)}\left( {P_3^{\left( i \right)} + R} \right) + P_4^{\left( i \right)}{C_i}}}} \right]
\end{array} \right..$$
Meanwhile, $k{x_i} + h{y_i} = \frac{{{D_i}}}{{{w_i}}}{C_i}$, then $\left( {{k_i}P_2^{\left( i \right)} + \frac{{{h_i}P_4^{\left( i \right)}{C_i}}}{{2R + P_5^{\left( i \right)}}}} \right)\left[ {\frac{R}{{P_2^{\left( i \right)}\left( {P_3^{\left( i \right)} + R} \right) + P_4^{\left( i \right)}{C_i}}}} \right] \approx \frac{{{D_i}}}{{{w_i}}}$, while $\frac{{{h_i}P_4^{\left( i \right)}{C_i}}}{{\left( {2R + P_5^{\left( i \right)}} \right){k_i}P_2^{\left( i \right)}}} \sim \frac{{{C_i}}}{R} \ll 1$. With all these approximation, then 

\begin{equation}
{C_1} = \frac{{P_2^{\left( 1 \right)}R\left( {{{{w_1}{k_1}} \mathord{\left/
{\vphantom {{{w_1}{k_1}} {{D_1}}}} \right.
\kern-\nulldelimiterspace} {{D_1}}} - 1} \right) - P_2^{\left( 1 \right)}P_3^{\left( 1 \right)}}}{{P_4^{\left( 1 \right)}}},
\label{Eq_S29}
\end{equation}

\begin{equation}
{C_2} = \frac{{P_2^{\left( 2 \right)}R\left( {{{{w_2}{k_2}} \mathord{\left/
{\vphantom {{{w_2}{k_2}} {{D_2}}}} \right.
\kern-\nulldelimiterspace} {{D_2}}} - 1} \right) - P_2^{\left( 2 \right)}P_3^{\left( 2 \right)}}}{{P_4^{\left( 2 \right)}}}.
\label{Eq_S30}
\end{equation}

We assume that the population dynamics of the resources follow Eq.9,then, for biotic resources, 
\begin{equation}
R = \frac{{\sqrt {{P_6}^2 + 4{P_6}{P_7}} - {P_6}}}{2},
\label{Eq_S31}
\end{equation}
where ${P_6} \equiv \frac{{{K_0}}}{{{r_0}}}[\frac{{P_2^{\left( 1 \right)}({k_1} - {{{D_1}} \mathord{\left/
{\vphantom {{{D_1}} {{w_1}}}} \right.
\kern-\nulldelimiterspace} {{w_1}}})}}{{P_4^{\left( 1 \right)}}} + \frac{{P_2^{\left( 2 \right)}({k_2} - {{{D_2}} \mathord{\left/
{\vphantom {{{D_2}} {{w_2}}}} \right.
\kern-\nulldelimiterspace} {{w_2}}})}}{{P_4^{\left( 2 \right)}}} - {r_0}]$ and
${P_7} = \frac{{{K_0}}}{{{r_0}}}(\frac{{{D_1}}}{{{w_1}}}\frac{{P_2^{\left( 1 \right)}P_3^{\left( 1 \right)}}}{{P_4^{\left( 1 \right)}}} + \frac{{{D_2}}}{{{w_2}}}\frac{{P_2^{\left( 2 \right)}P_3^{\left( 2 \right)}}}{{P_4^{\left( 2 \right)}}})$ .
For abiotic resources,
\begin{equation}
R = \frac{{{r_0} + (\frac{{{D_1}}}{{{w_1}}}\frac{{P_2^{\left( 1 \right)}P_3^{\left( 1 \right)}}}{{P_4^{\left( 1 \right)}}} + \frac{{{D_2}}}{{{w_2}}}\frac{{P_2^{\left( 2 \right)}P_3^{\left( 2 \right)}}}{{P_4^{\left( 2 \right)}}})}}{{\frac{{P_2^{\left( 1 \right)}(k - {D_1}/{w_1})}}{{P_4^{\left( 1 \right)}}} + \frac{{P_2^{\left( 2 \right)}(k - {D_2}/{w_2})}}{{P_4^{\left( 2 \right)}}} + \frac{{{r_0}}}{{{K_0}}}}}.
\label{Eq_S32}
\end{equation} 
Eqs.\ref{Eq_S29}-\ref{Eq_S32} are the analytical solutions to the steady-state species abundances under the approximation that $R \gg {C_1},{C_2}$. The analytical solutions are compared to that of numerical results (exact solutions) in Fig.3f and Fig.4, which shows good agreement.

\section{ Breaking CEP for any number of resource species.}
We have already illustrated that in case $N=1$ and $M=2$, both species of consumers can coexist at steady state and thus break the constraint of the CEP (Fig.2d and f). Here we show that for any $N>0$, the constraint of CEP can be liberated. When $N \ge 2$, we construct the following scenario that $M = N + 1$ species of consumers can coexist at steady state in a natural ecosystem: For consumer species $C_i$ ($i$=1-$N-1$), each species only feeds on one resource species $R_i$ ($i$=1-$N-1$), respectively. Meanwhile, consumer species $C_N$ and $C_{N+1}$ only feed on $R_N$. Then, similar to the case of $N=1$ and $M=2$, $C_N$ and $C_{N+1}$ can coexist. Meanwhile, similar to the case in SI Sec.I.A.2, species $C_i$ ($i$=1-$N-1$) can coexist together with $C_N$ and $C_{N+1}$. Consequently, all $N+1$ species of consumers can coexist at steady state, with $M=N+1>N$.

\section{Dimensional analysis for Models involving chasing triplet.}
The equations for the population dynamics of Model A are shown in Eq.8 and Eq.\ref {Eq_S17}. For biotic resource cases:
\begin{equation}
\left\{ \begin{array}{l}
\dot{x_1}={a_1}{R^{\left( {\rm{F}} \right)}}C_1^{\left( {\rm{F}} \right)} - \left( {{d_1} + {k_1}} \right){x_1} - {b_1}{x_1}C_1^{\left( {\rm{F}} \right)} + {e_1}{y_1}\\
\dot{x_2}={a_2}{R^{\left( {\rm{F}} \right)}}C_2^{\left( {\rm{F}} \right)} - \left( {{d_2} + {k_2}} \right){x_2} - {b_2}{x_2}C_2^{\left( {\rm{F}} \right)} + {e_2}{y_2}\\
\dot{y_1}={b_1}{x_1}C_1^{\left( {\rm{F}} \right)} - \left( {{h_1} + {e_1} + {l_1}} \right){y_1}\\
\dot{y_2}={b_2}{x_2}C_2^{\left( {\rm{F}} \right)} - \left( {{h_2} + {e_2} + {l_2}} \right){y_2}\\
\dot{C_1}= {w_1}\left( {{k_1}{x_1} + {h_1}{y_1}} \right) - {D_1}{C_1}\\
\dot{C_2}= {w_2}\left( {{k_2}{x_2} + {h_2}{y_2}} \right) - {D_2}{C_2}\\
\dot{R}= R{R_0}\left( {1 - {R \mathord{\left/
{\vphantom {R {{K_0}}}} \right.
\kern-\nulldelimiterspace} {{K_0}}}} \right) - \left( {{k_1}{x_1} + {h_1}{y_1}} \right) - \left( {{k_2}{x_2} + {h_2}{y_2}} \right)
\end{array} \right.
\label{Eq_S33}
\end{equation} 
Define dimentionless variables $T,{X_1},{X_2},{Y_1},{Y_2},C_1^\prime ,C_2^\prime ,R',C_i^{\left( {\rm{F}} \right)\left( {\dim } \right)},R^{\left( {\rm{F}} \right)\left( {\dim } \right)}$ as follows.
\begin{equation}
\left\{ \begin{array}{l}
T \equiv t/{T_0};\\
{X_1} \equiv {x_1}/{x_{10}},{X_2} \equiv {x_2}/{x_{20}};\\
{Y_1} \equiv {y_1}/{y_{10}},{Y_2} \equiv {y_2}/{y_{20}};\\
C_1^\prime \equiv {C_1}/{C_{10}},C_2^\prime \equiv {C_1}/{C_{20}},R' \equiv R/{r_0};\\
C_i^{\left( {\rm{F}} \right)\left( {\dim } \right)} \equiv C_i^\prime - {X_i} - 2{Y_i},\quad i = 1,2;\\
{R^{\left( {\rm{F}} \right)\left( {\dim } \right)}} \equiv R' - \left( {{X_1} + {X_2} + {Y_1} + {Y_2}} \right),
\end{array} \right.
\label{Eq_S34}
\end{equation} 
and we define dimensionless parameters (marked with `$\equiv$') and chose the flexibe parameters as follow 
\begin{equation}
\left\{ \begin{array}{l}
{T_0} = {N_1}{\left( {{D_2}} \right)^{ - 1}},{R_0}' \equiv {R_0}{T_0},{D_1}' \equiv {D_1}{T_0},{D_2}' \equiv {D_2}{T_0} = {N_1};\\
{e_i}' \equiv {e_i}{T_0},{l_i}' \equiv {l_i}{T_0},{k_i}' = {k_i}{T_0},{h_i}' = {h_i}{T_0},{d_i}' = {d_i}{T_0},\quad i = 1,2;\\
{a_1}' \equiv {a_1}{T_0}{x_{10}},{a_2}' \equiv {a_2}{T_0}{x_{10}},{b_1}' \equiv {b_1}{T_0}{x_{10}},{a_2}' \equiv {a_2}{T_0}{x_{10}};\\
{r_0} = {C_{10}} = {C_{20}} = {x_{10}} = {x_{20}} = {y_{10}} = {y_{20}} = {K_0}/{N_2}.
\end{array} \right.
\label{Eq_S35}
\end{equation} 
Here, $N_1$ and $N_2$ are two reducible parameters which can be either 1 or arbitrary positive numbers. Substituting Eqs.\ref{Eq_S34}-.\ref{Eq_S35} into Eq.\ref{Eq_S33}, we get
\begin{equation}
\left\{ \begin{array}{l}
\dot{X_i}={a_i}'{R^{\left( {\rm{F}} \right)\left( {\dim } \right)}}C_i^{\left( {\rm{F}} \right)\left( {\dim } \right)} - \left( {{d_i}' + {k_i}'} \right){X_i} - {b_i}'{X_i} C_i^{\left( {\rm{F}} \right)\left( {\dim } \right)} + {e_i}'{Y_i}\\
\dot{Y_i}={b_i}' {X_i}C_i^{\left( {\rm{F}} \right)\left( {\dim } \right)} - \left( {{h_i}' + {e_i}' + {l_i}'} \right) {Y_i}\\
\dot{{C_1}'} = {w_1}\left( {{k_1}' {X_1} + {h_1}'{Y_1}} \right) - {D_1}' {C_1}'\\
\dot{{C_2}'} = {w_2}\left( {{k_2}' {X_2} + {h_2}'{Y_2}} \right) - {N_1} {C_2}'\\
\dot{R'} = R' \cdot {R_0}'(1 - R'/{N_2}) - ({k_1}'{X_1} + {h_1}'{Y_1}) - ( {k_2}'{X_2}+ {h_2}'{Y_2})
\end{array} \right..
\label{Eq_S36}
\end{equation} 
Note that all variables and parameters in Eq.\ref{Eq_S36} are dimensionless. Compare Eq.\ref{Eq_S36} with Eq.\ref{Eq_S33}, it is clear that all equations have the same form except that $N_1$ and $N_2$ in Eq.\ref{Eq_S36} are reducible which can be either 1 or arbitrary positive numbers.
Similarly, for the abiotic resource case in Model A, or the biotic/abiotic resource cases in Model B and Model C, only two parameters: $D_2$ and $K_0$ are reducible in the dimensionless expressions, which can be set as either 1 or arbitrary positive numbers. For convenience, in our numerical calculations, we use the same parameter notation while they are all dimensionless parameters. For the choice of $N_1$ ($D_2$) and $N_2$ ($K_0$), in the biotic resource case, we set $D_2$=0.005, $K_0$=10; in the abiotic resource case, we set $D_2$=0.004, $K_0$=5.



\begin{figure}
\includegraphics[width=0.7\linewidth]{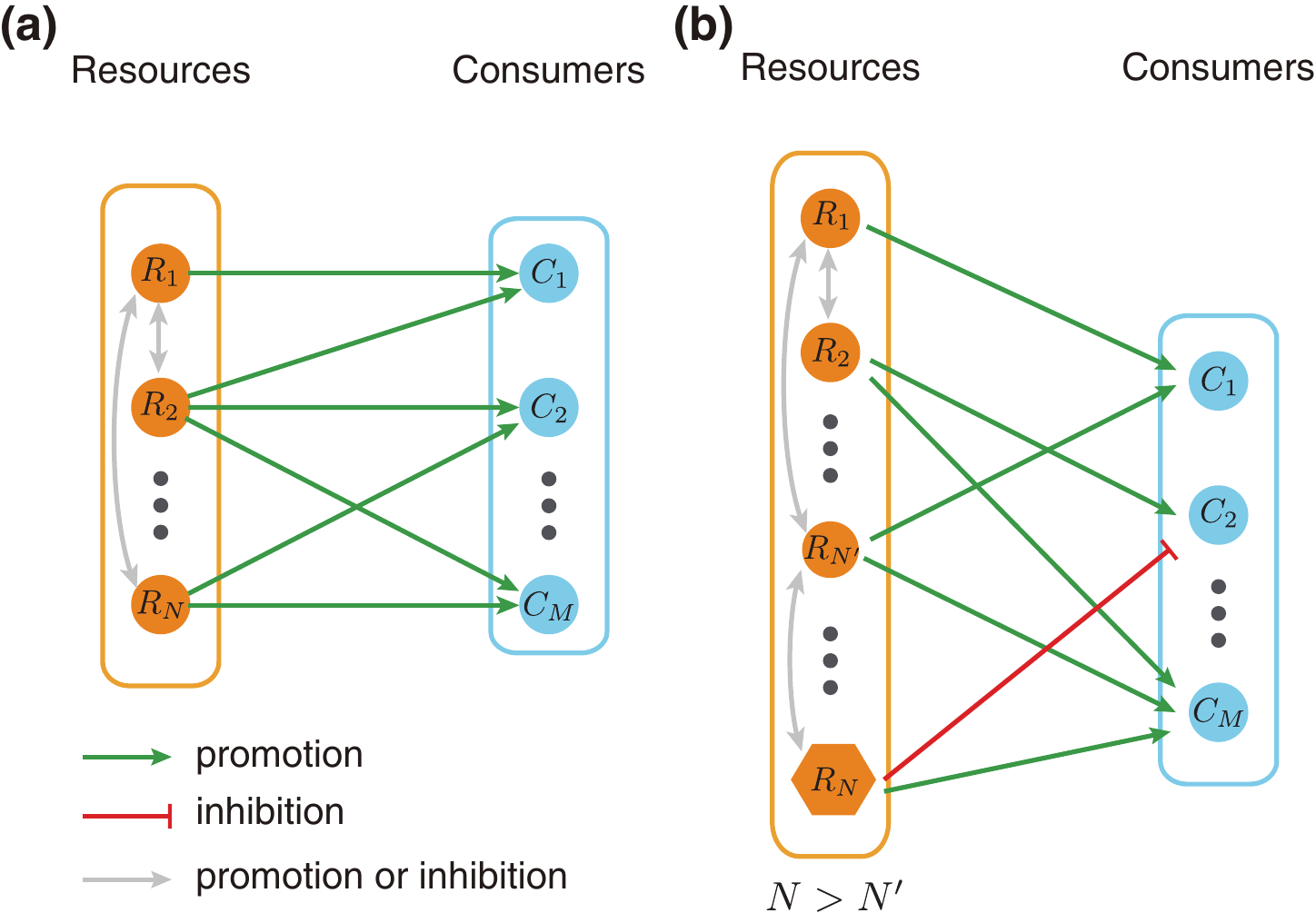}%
\caption{ Bipartite graph between resources and consumers. (a) $M$ species of consumers feed on $N$ species of resources. Predation or other interactions are forbidden among consumers but allowed among resources. Competitive exclusion principle (CEP) states that at steady state the coexisting $M \le N$. (b) Resources involve chemical compounds: $R_i$ ($N' + 1 \le i \le N$; $N > N'$) are chemical compounds, which can promote or inhibit the growth of consumers, while $R_i$ ($1 \le i \le N'$) are normal resources, supplying as food for consumers. In total, there are $N$ species of resources and $M$ species of consumers. According to CEP, at steady state, it is permitted that the coexisting $M>N'$, yet $M \le N$.}
\label {FigS1}
\end{figure}

\begin{figure}
\includegraphics[width=0.5\linewidth]{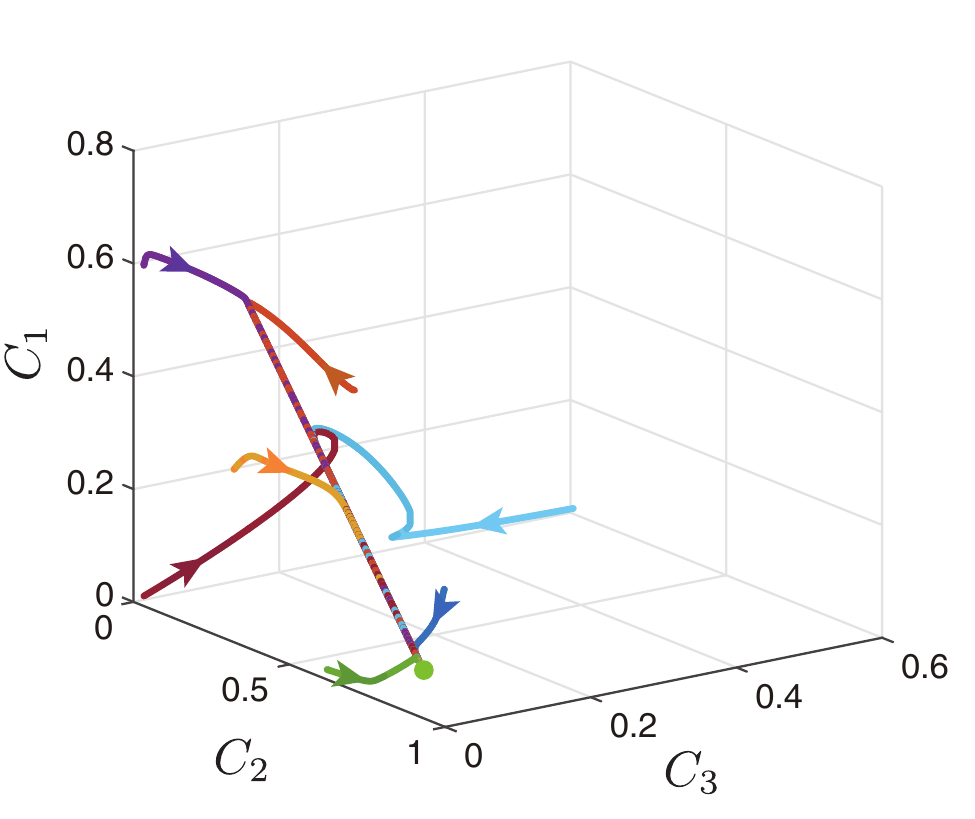}%
\caption{ Representative phase portrait of the trajectories following the classical proof. $M=3$ and $N=2$. Two consumer species, at most, can coexist at steady state. Here equations are that shown in Eq.2 where ${f_i}\left( {{R_1},{R_2}} \right) = \alpha _i^{\left( 1 \right)}{R_1} + \alpha _i^{\left( 2 \right)}{R_2}$ ($i$ =1, 2, 3); ${g_j}\left({{R_1},{R_2},{C_1},{C_2},{C_3}} \right) = {R_j}\left( {r_j^{\left( 0
\right)} - \beta _j^{\left( 0 \right)} {R_j} - \beta _j^{\left( 1 \right)}{C_1} - \beta
_j^{\left( 2 \right)}{C_2} - \beta _j^{\left( 3 \right)}{C_3}}
\right)$ ($j$=1, 2); ${D_1} = 0.0006$; ${D_2} = 0.0005$; ${D_3} = 0.0004$; $\alpha _1^{\left( 1 \right)} = 0.0013$; $\alpha _2^{\left( 1 \right)} = 0.0011$; $\alpha _1^{\left( 2 \right)} = 0.001$; $\alpha _2^{\left( 2 \right)} = 0.0009$; $r_1^{\left( 0 \right)} = 1.01$; $r_2^{\left( 0 \right)} = 1$; $\beta _1^{\left( 0 \right)}=\beta _2^{\left( 0 \right)} = 1$, $\beta _1^{\left( 1 \right)} = 1.3$; $\beta _2^{\left( 1 \right)} = 1$; $\beta _1^{\left( 2 \right)} = 1.1$; $\beta _2^{\left( 2 \right)} = 0.9$; $\alpha _3^{\left( 1 \right)}=0.0001$; $\alpha _3^{\left( 2 \right)}=0.0021$; $\beta _1^{\left( 3 \right)} = 0.1$ and $\beta _2^{\left( 3 \right)} = 2.1$. In the initial condition, ${R_1} = 0.01$ and ${R_2} = 0.01$ for all trajectories. Finally, all trajectories converge at the green point with $C_1=0$.}
\label {FigS2}
\end{figure}

\begin{figure}
\includegraphics[width=0.65\linewidth]{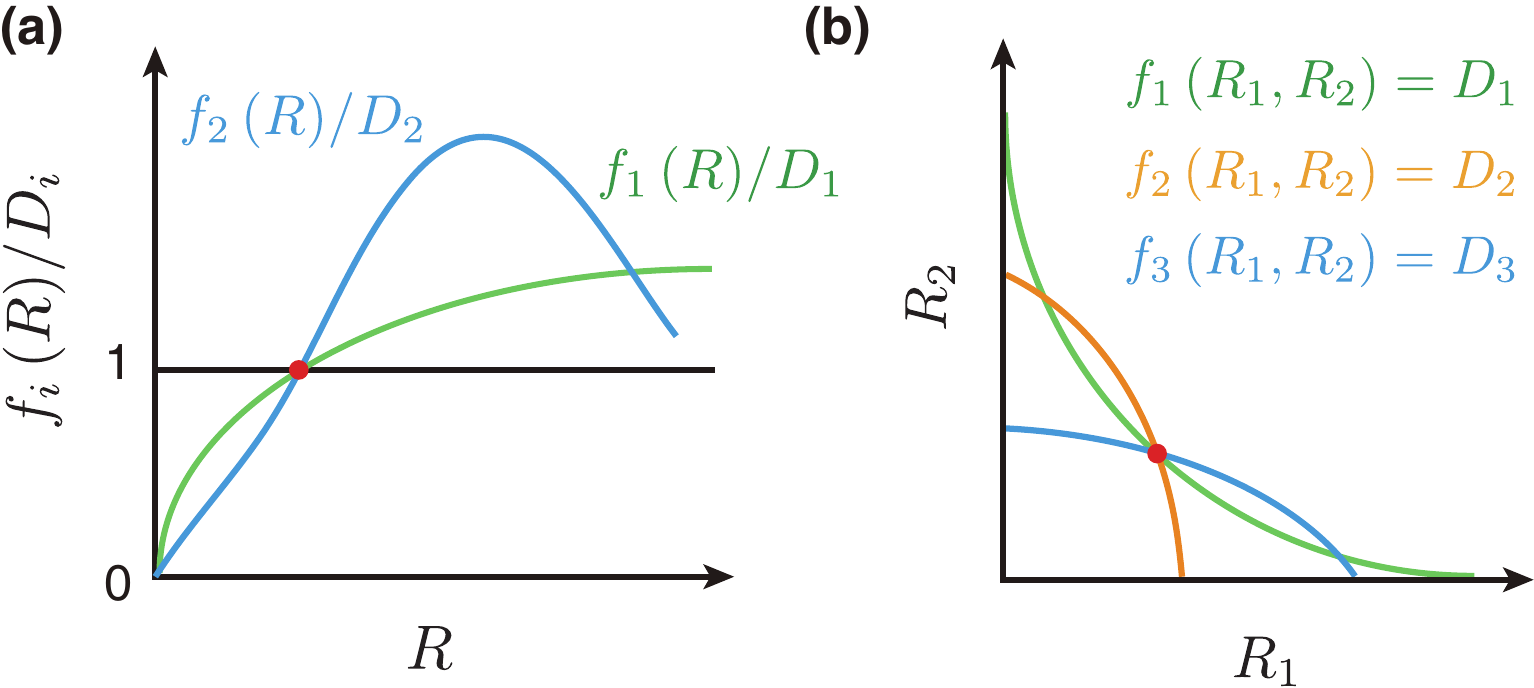}%
\caption{ Special cases permit $M>N$ at steady state. (a) $M=2$ and $N=1$. (b) $M=3$ and $N=2$. Within the classical CEP framework, it is possible that $M>N$ at steady state for a special parameter set (with Lebesgue measure zero), which corresponds to the case that three lines accidentally intersect at a common point (red point in a and b). A simple scheme of this special case is $f_i=f$, $D_i=D$ ($i$=1-$M$).}
\label {FigS3}
\end{figure}

\begin{figure}
\includegraphics[width=0.7\linewidth]{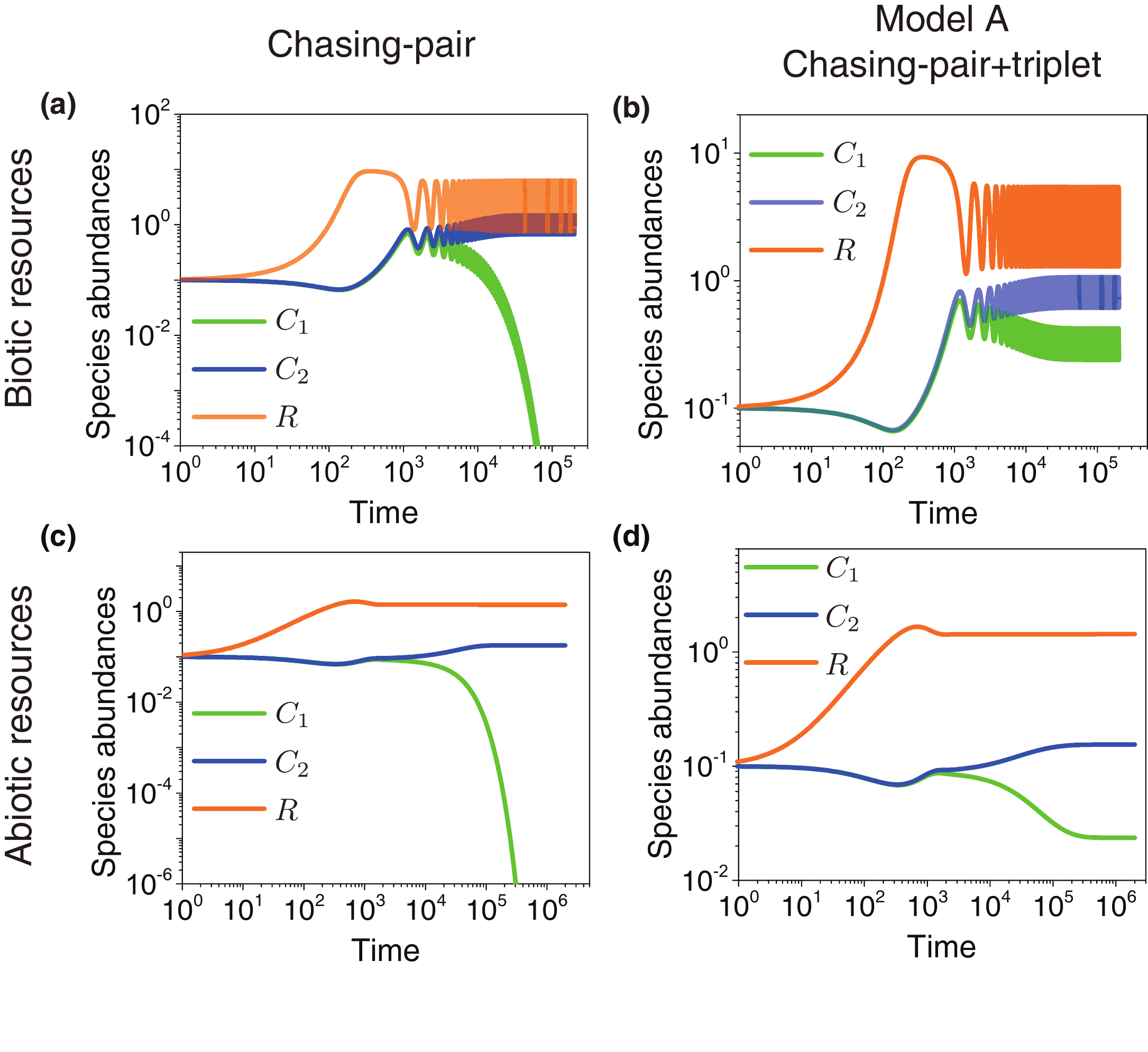}%
\caption{Time course of the abundance of two consumer species ($M$=2) and one resource species ($N$=1). (a) In the presence of only chasing pairs, only one type of consumer species exists for long, the oscillating dynamics resembles that of the classical predator-prey models ~\cite{RN526}. (b) In Model A (the presence of chasing pair and triplet), two consumer species coexist at oscillating abundances. (c) In the presence of only chasing pairs, consumer species cannot coexist at steady state. (d) In Model A, both consumer species coexist at steady state. (a)-(b) Biotic resources, (c)-(d) Abiotic resources.
(a) was simulated from Eq.5, where $g = R{R_0}(1 - R/{K_0}) - ({k_1}{x_1} + {k_2}{x_2})$; (b) was simulated from Eqs.8-9; 
(c) was simulated from Eq.5, where $g = {R_0}(1 - R/{K_0}) - ({k_1}{x_1} + {k_2}{x_2})$; (d) was simulated from Eqs.8-9.
In (a)-(d): $a_i=0.1$, $d_i=0.1$, $k_i=0.1$, $w_i=0.1$ ($i$=1, 2); the initial abundances of $\left( {R,{C_1},{C_2}} \right)$ are $(0.1,0.1, 0.1)$.
In (a)-(b): $D_2=0.005$, $K_0=10$, $R_0=0.03$, $D_1=1.03D_2$.
In (c)-(d): $D_2=0.004$, $K_0=5$, $R_0=0.01$, $D_1=1.01D_2$ . 
In (b) and (d): $b_i=0.1$, $e_i=0.1$, $h_i=0.1$, $l_i=0.1$ ($i$=1, 2).
}
\label {FigS4}
\end{figure}

\begin{figure}
\includegraphics[width=0.7\linewidth]{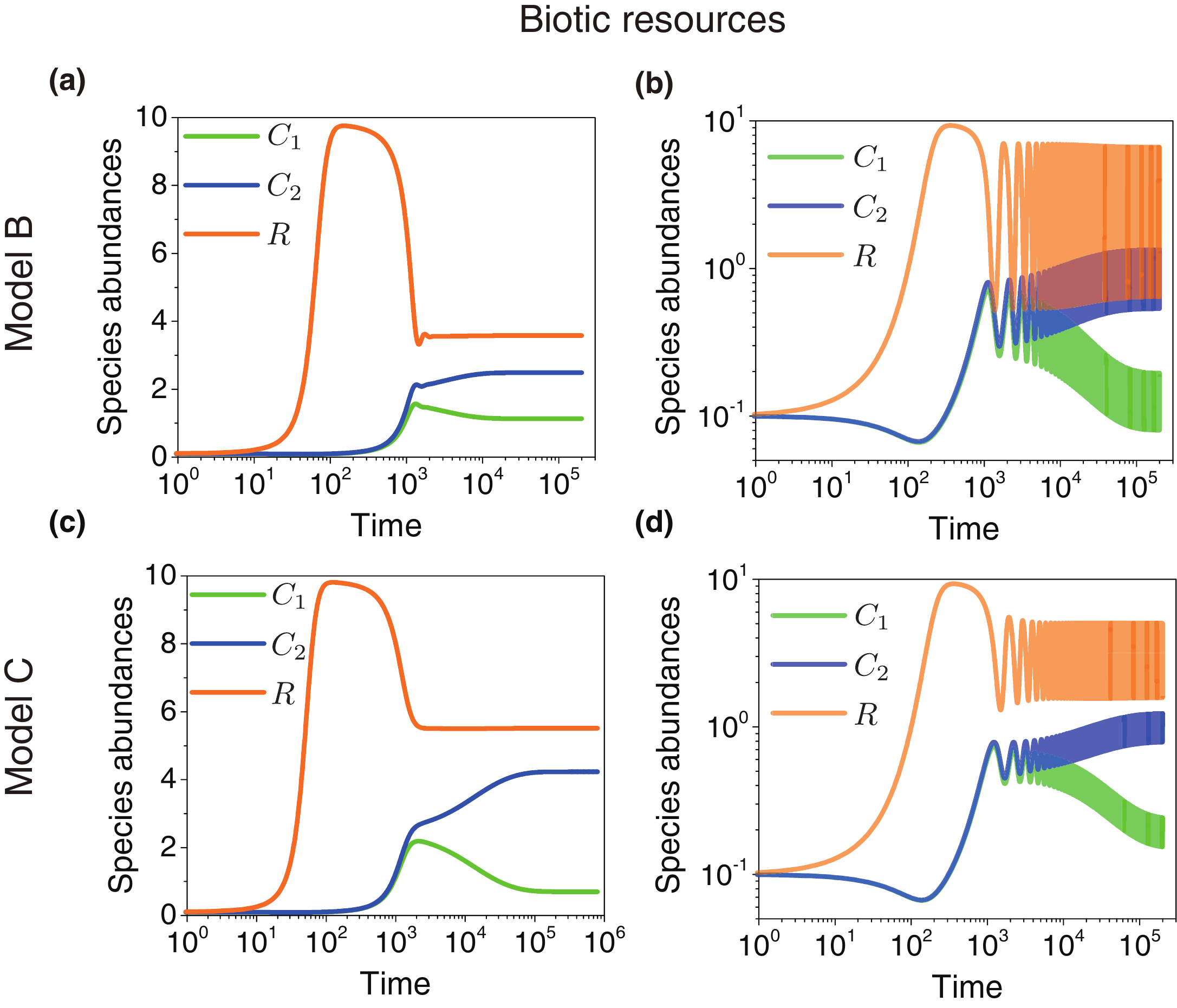}%
\caption{Time course of the abundance of two consumer species ($M$=2) and one biotic resource species ($N$=1). In the presence of chasing pair and chasing triplet, two consumer species coexist at steady state ((a), (c)) or with oscillating abundances ((b), (d)). (a)-(b) Model B. (c)-(d) Model C.
(a)-(b) were simulated from Eqs.16-17; 
(c)-(d) were simulated from Eqs.18-19; 
In (a)-(d): $a_i=p_i=0.1$, $d_i=s_i=t=0.1$, $k_i=0.1$, $w_i=0.1$ ($i$=1, 2); $D_2=0.005$, $K_0=10$, the initial abundances of $\left( {R,{C_1},{C_2}} \right)$ are $(0.1,0.1, 0.1)$.
In (a): $R_0=0.08$, $D_1=1.05D_2$, $q_i=0.1$ ($i$=1, 2);
In (b): $R_0=0.03$, $D_1=1.02D_2$, $q_i=0.1$ ($i$=1, 2);
In (c)-(d): $q_i=0.05$, $b_i=0.1$, $e_i=l_i=0.1$, $h_i=0.1$ ($i$=1, 2); 
In (c): $R_0=0.1$, $D_1=1.02D_2$;
In (d): $R_0=0.03$, $D_1=1.01D_2$.
}
\label {FigS5}
\end{figure}

\begin{figure}
\includegraphics[width=0.6\linewidth]{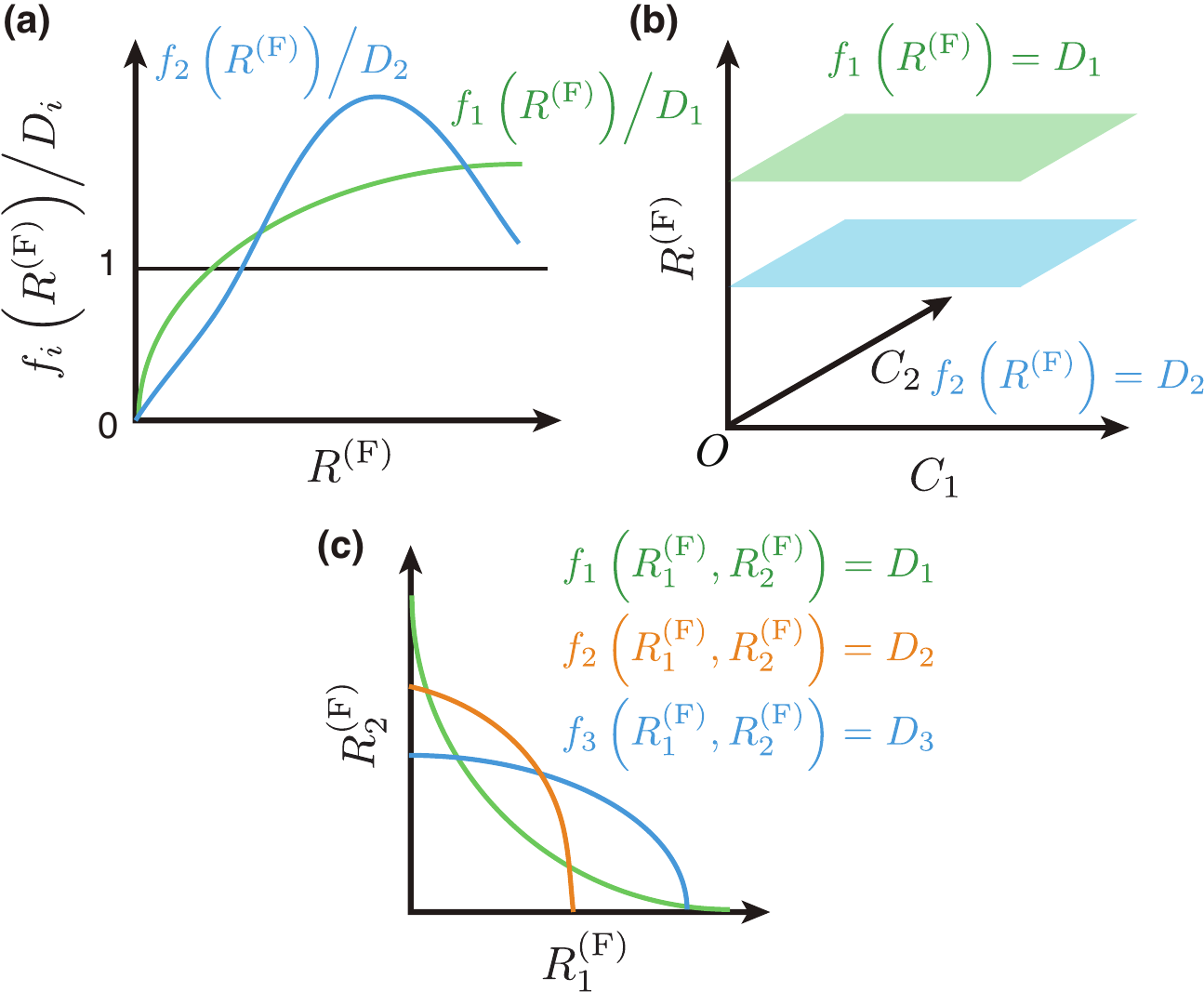}%
\caption{Chasing-pair scenarios are still under the constraint of competitive exclusion. (a) $M=2$ and $N=1$. If all consumer species coexist at steady state, ${f_i}\left( {{R^{\left( {\rm{F}} \right)}}} \right)/{D_i} = 1$ ($i$=1, 2). This requires that three lines $y = {f_i}\left( R \right)/{D_i}$ ($i$=1, 2) and $y=1$ share a common point, which normally cannot happen. (b) $M=2$ and $N=1$. The green plane is parallel to the blue one, and hence they do not have a common point. (c) $M=3$ and $N=2$. At steady state, if all consumer species coexist, then ${f_i}\left( {R_1^{\left( {\rm{F}} \right)},R_2^{\left( {\rm{F}} \right)}} \right) = {D_i}$ ($i$=1-3). But three lines normally do not intersect at a common point.}
\label {FigS6}
\end{figure}

\begin{figure}
\includegraphics[width=0.55\linewidth]{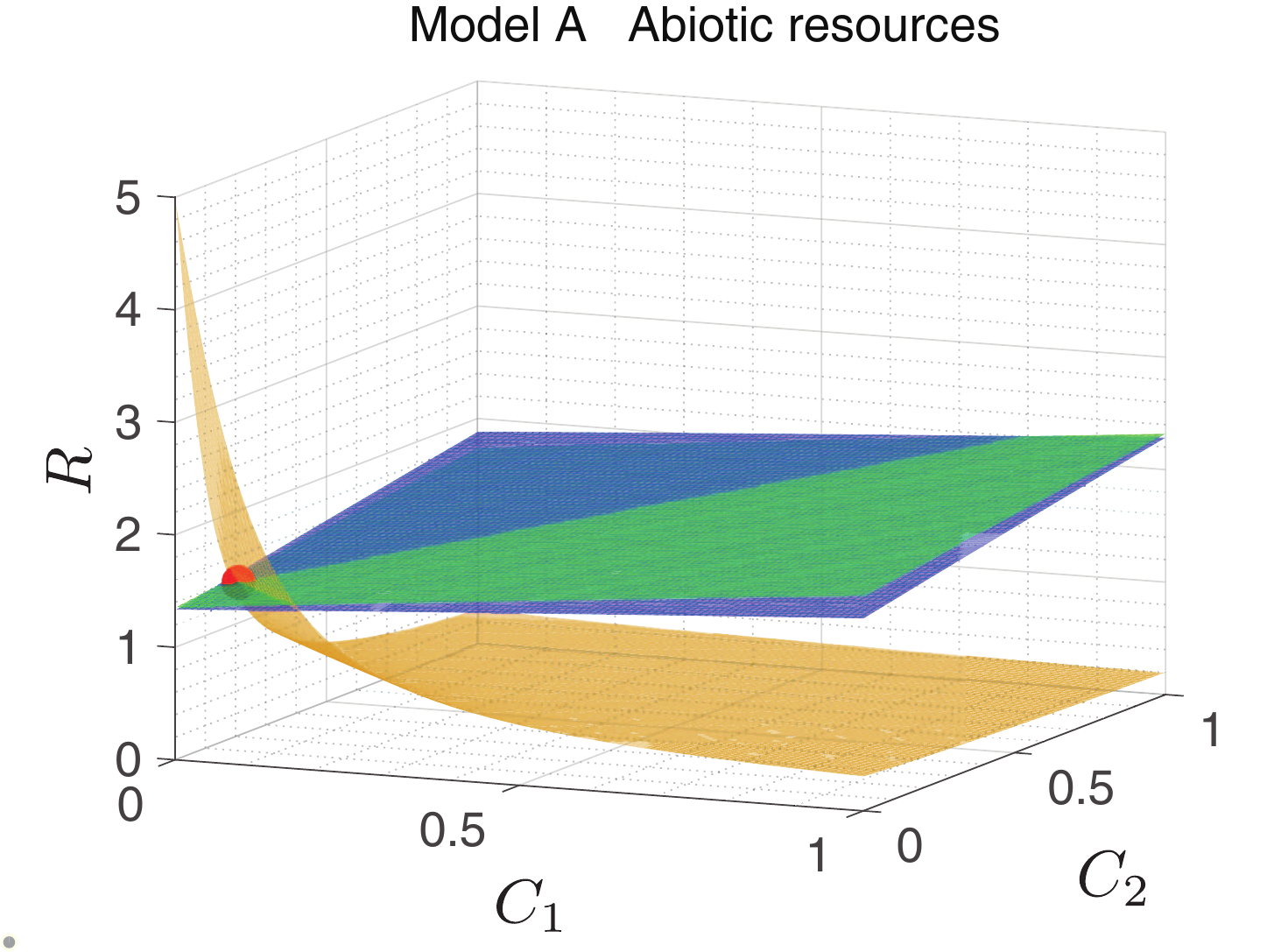}%
\caption{Demonstration of the intuitive explanation with numerical solutions. In Model A (with chasing pair and triplet) of abiotic resources, numerical solutions confirm that the yellow, green and blue surfaces are not parallel to each other and definitely can have a common point (marked with red dot).
This figure was calculated using Eqs.8-9.
Here, $a_i=b_i=0.1$, $d_i=e_i=l_i=0.1$, $k_i=h_i=0.1$, $w_i=0.1$ ($i$=1, 2); $D_2=0.004$, $K_0=5$, $R_0=0.01$, $D_1=1.01D_2$ . 
}
\label {FigS7}
\end{figure}

\begin{figure}
\includegraphics[width=0.7\linewidth]{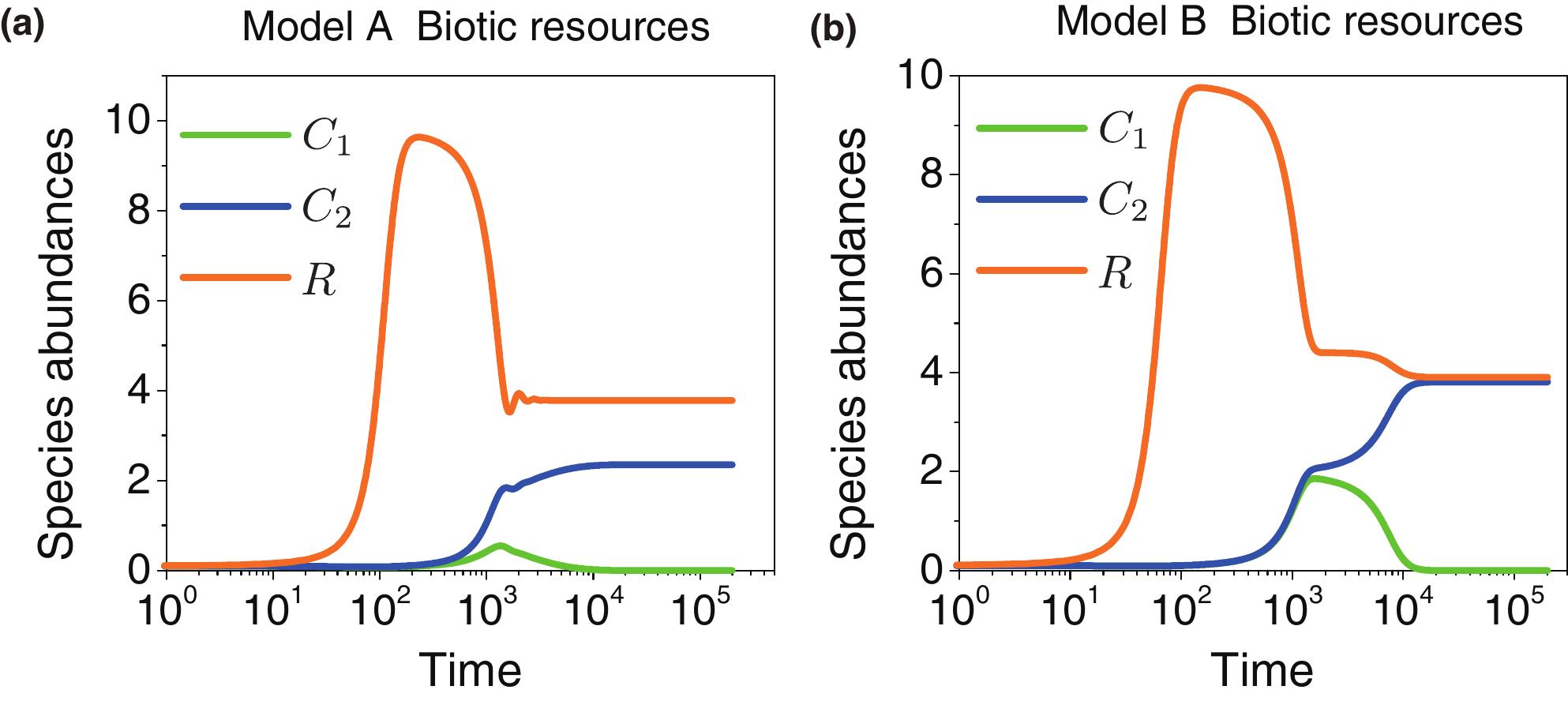}%
\caption{Time course of the abundance of two consumer species ($M$=2) and one biotic resource species ($N$=1). Forming chasing triplet is not a guarantee for breaking CEP. (a) Time series profile corresponds to Fig.8d. (b) Time series profile corresponds to Fig.8f.
(a) was calculated using Eqs.8-9; (b) was calculated using Eqs.16-17.
In (a)-(b): $a_i=0.1$, $d_i=0.1$, $k_i=0.1$, $w_i=0.1$ ($i$=1, 2); $D_2=0.005$, $K_0=10$, the initial abundances of $\left( {R,{C_1},{C_2}} \right)$ are $(0.1,0.1, 0.1)$.
In (a): $b_i=0.1$, $e_i=0.1$, $h_i=0.1$, $l_i=0.1$ ($i$=1, 2); $R_0=0.05$, $D_1=1.2D_2$. 
In (b): $p_i=0.1$, $s_i=0.1$, $q_i=0.05$ ($i$=1, 2); $t=0.1$, $R_0=0.08$, $D_1=1.01D_2$.
}
\label {FigS8}

\end{figure}


\end{widetext}
\end{document}